\documentclass[manuscript,nonacm]{acmart}

\usepackage{appendix}
\usepackage{booktabs}
\usepackage{colortbl}
\usepackage{float}
\usepackage{graphicx}
\usepackage{multirow}
\usepackage{subfig}
\usepackage{tabularx}
\usepackage{textcomp}
\usepackage{url}

\begin{document}

\title{Security Best Practices: A Critical Analysis Using IoT as a Case Study}

\author{David Barrera}
\affiliation{%
    \institution{Carleton University}
    \city{Ottawa}
    \state{Ontario}
    \country{Canada}
}

\author{Christopher Bellman}
\affiliation{%
    \institution{Carleton University}
    \city{Ottawa}
    \state{Ontario}
    \country{Canada}
}
\authornote{Authors listed in alphabetical order. Contact author: Christopher Bellman (chris@ccsl.carleton.ca). A version of this paper is to appear in ACM TOPS. A preliminary version of part of this work appeared previously as a tech report \cite{Bellman2020}.}

\author{Paul C. van Oorschot}
\affiliation{%
    \institution{Carleton University}
    \city{Ottawa}
    \state{Ontario}
    \country{Canada}
}

\renewcommand{\shortauthors}{Barrera, Bellman, and van Oorschot}

\begin{abstract}
    Academic research has highlighted the failure of many Internet of Things (IoT) product manufacturers to follow accepted practices, while IoT security \textit{best practices} have recently attracted considerable attention worldwide from industry and governments. Given current examples of security advice, confusion is evident from guidelines that conflate desired outcomes with security practices to achieve those outcomes. We explore a surprising lack of clarity, and void in the literature, on what (generically) \textit{best practice} means, independent of identifying specific individual practices or highlighting failure to follow best practices. We consider categories of security advice, and analyze how they apply over the lifecycle of IoT devices. For concreteness in discussion, we use iterative inductive coding to code and systematically analyze a set of 1013 IoT security best practices, recommendations, and guidelines collated from industrial, government, and academic sources. Among our findings, of all analyzed items, 68\% fail to meet our definition of an (actionable) practice, and 73\% of all actionable advice relates to the software development lifecycle phase, highlighting the critical position of manufacturers and developers. We hope that our work provides a basis for the community to better understand best practices, identify and reach consensus on specific practices, and find ways to motivate relevant stakeholders to follow them.
\end{abstract}


\maketitle

\section{Introduction}
\label{sec:introduction}
Internet of Things (IoT) is commonly described as adding network connectivity to traditionally non-networked items or ``things'' \cite{Wortmann2015}. It surrounds us with a variety of network-connected devices such as smart light bulbs, door locks, web cameras, audio speakers, thermostats, and less obvious objects like fridges, traffic lights, or sensors and controllers built into critical infrastructure systems. The importance of IoT in marketing and sales has resulted in a wide variety of devices with arguably unnecessary functionality (e.g., internet-connected toasters and toys). These devices, while offering convenience or new functionality, have acquired a reputation \cite{Alrawi2019} of poor security and misconfiguration, leading to huge numbers of network-accessible devices being vulnerable. As IoT devices may be more isolated or resource-constrained (e.g., battery power, processors, memory) than their Internet of Computers (IoC---i.e., pre-IoT devices such as mobile phones, laptop/desktop computers, servers) counterparts, or lacking in software update support, their security issues are often hard to address. The cyberphysical nature of IoT---interfacing with physical world objects---also results in threats to our physical world, as well as to networks and other internet hosts \cite{Kolias2017}. This has resulted in considerable attention (e.g., \cite{Alrawi2019,ETSI2020,IEEE1,Microsoft1}) to best practices for IoT security.

The term \textit{best practice} is commonly assumed to be intuitively understood, yet academic work in this area (as noted below) lacks consensus on informal definitions for the term, and closer inspection suggests a clear explicit definition is needed. We argue that this assumption results in ambiguity, and contributes to security problems, and that intuitive understandings are at best foggy and differ considerably across even experts. For example, in considering Cloud Security Providers (CSPs), Huang et al. \cite{huang2015} refer to: ``\textit{security mechanisms that have been implemented across a large portion of the CSP industry [are thus] considered standardized into a `best-practice'.}'' Here, best practice appears to mean \textit{widely implemented}. In their evaluation of home-based IoT devices, Alrawi et al. \cite{Alrawi2019} note numerous violations of security design principles, and assert ``\textit{Best practices and guidelines for the IoT components are readily available}'', but offer neither citations for best practices among 108 references, nor their own definition. In a recent national news article \cite{CBC2020} on banks disclaiming liability for customer losses from e-transfer fraud, and one-sided online banking agreements, a defensive bank representative is quoted: ``\textit{We regularly review our policies and procedures to ensure they align with best practices.}'' This quote appears to be not about security, but rather legal best practices in the sense of \textit{our agreements are no worse than our competitors'}. Large collections of documents from industrial, government, and academic sources also conflate best practice with common terms such as \textit{recommendation} and \textit{guideline} \cite{IoTSecMap}. How do best practices, good practices, and standard practices differ? Or guidelines, recommendations, and requirements? If something is not \textit{actionable}, does it make sense to recommend it as a best practice? 

In this paper, we provide what we believe is the first in-depth technical examination of intended meanings of the term \textit{security best practice}, and the surrounding related terms noted above. We argue that confusion and ambiguity result from the lack of a common understanding and precise definition of these terms, and that this confusion permeates official best practice recommendations. We support this argument by first investigating current use of terms related to best practices, and explain how meanings of each term differ qualitatively (Section~\ref{sec:bestpractices}). We classify these descriptive terms into three categories and separately define (actionable) security \textit{practices} distinct from \textit{desired security outcomes} and security \textit{principles}. Our examination of terminology highlights ambiguity and conflation of established terms, contributing to the challenges we uncover in current IoT security advice and technical literature. 

As further contributions, we offer uniform, consistent terminology, and then consider the UK government's \textit{Code of Practice for Consumer IoT Security} \cite{DCMS2}. A preliminary informal analysis finds that the guidelines comprising it are not actionable practices by our definition (Section~\ref{sec:bestpractices/revisiting}), despite being positioned as ``practical steps'' to be taken by IoT security stakeholders, featuring 13 ``outcome-focused guidelines'' derived from industry advice. We develop a new security advice coding methodology (\textit{SAcoding method}) in Section~\ref{sec:methodology}, for systematically categorizing security advice based in part on the terminology refined herein. Applying it to a dataset of 1013 IoT security advice items from industrial, government, and academic sources compiled by the UK government \cite{DCMS1}, we find only 32\% of the 1013 items are actionable (Section~\ref{sec:analysis}), highlighting a gap between the expectations of entities providing advice and those intended to implement it. Our goal is not to criticise the UK group---their advice dataset simply aggregates other sources---but to demonstrate what we view as the ineffective state of IoT security advice as a whole, and to take first steps to repair this. We believe that our contributions may be of use to the broader security community, beyond IoT itself.

\section{Background and Overview of Established IoT Security Advice}
\label{sec:background}
In this section we discuss key areas of IoT and their role in the adoption of security best practices. These areas include: the IoT device lifecycle (and when in a device's lifecycle security advice is applicable), which stakeholders have the most significant impact on the security of a device, and existing security advice, which we later analyze. 

    \subsection{Lifecycle of IoT Devices}
    \label{sec:background/lifecycle}    
    
    The lifecycle of a consumer IoT device includes phases it goes through from early design to the time it is discarded (possibly re-used, or never used again) \cite{Garcia-Morchon2019}. We model the full lifecycle of a device, as decisions made within one part of the lifecycle (particularly the pre-deployment stages) may affect later phases. 
    
\begin{figure}[h]
        \centering
        \includegraphics[width=0.55\textwidth]{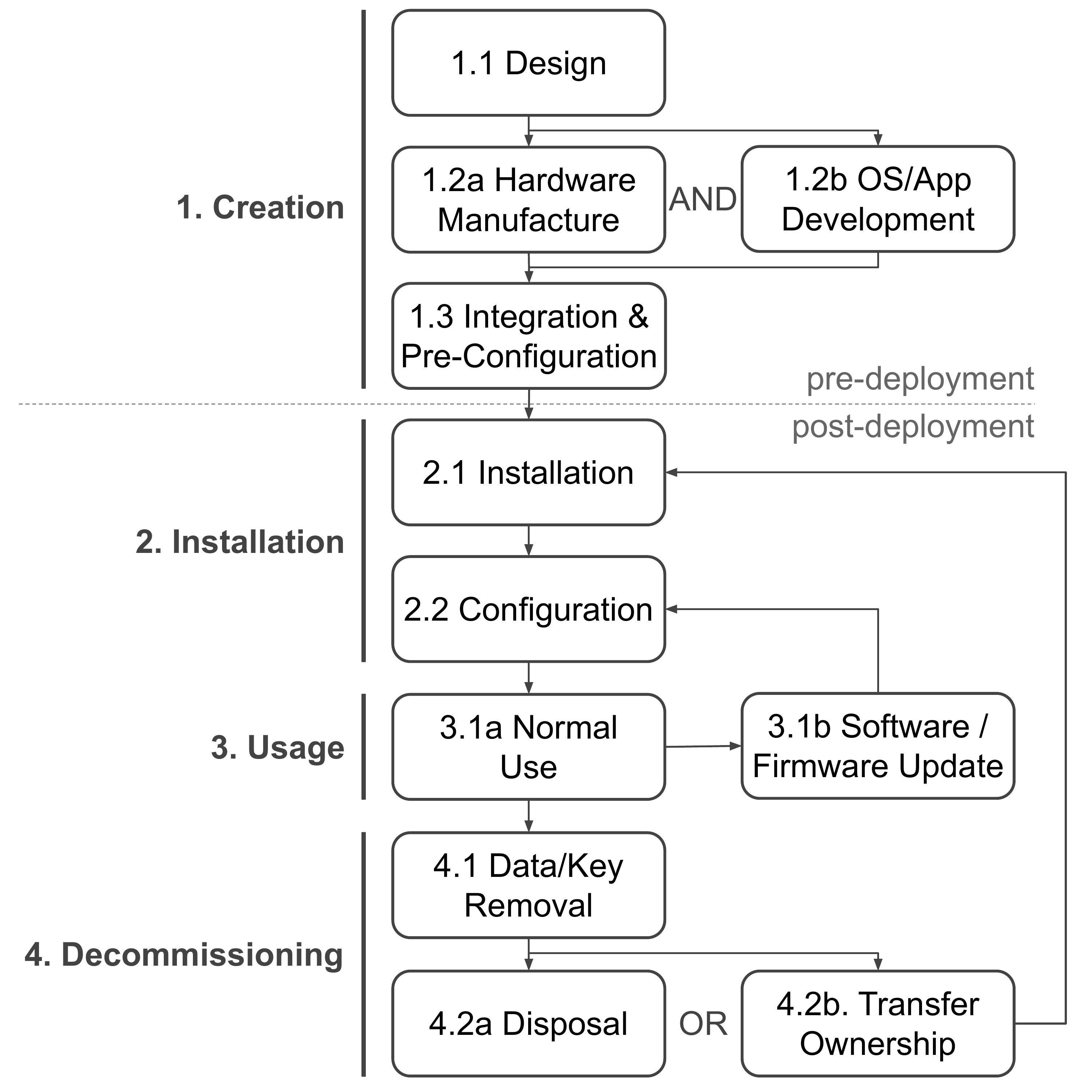}
        \caption{Typical IoT device lifecycle (initial design to end-of-life). Broad \textit{phases} (1--4) encapsulate multiple \textit{stages} (e.g., 2.1, 2.2).}
        \label{fig:lifecycle}
    \end{figure}

    Once IoT products have left the hands of manufacturers, it becomes more challenging to address vulnerabilities. We believe it is important to understand the stages within each phase, as security advice to be followed relates directly to the processes carried out within specific stages. Fig.~\ref{fig:lifecycle} presents our model of a typical lifecycle of an IoT device based on existing work \cite{Garcia-Morchon2019}, modified to incorporate what we believe are the most relevant phases (and stages within them). Our model highlights our interpretation of the four major phases where IoT security advice is generally applicable. Our analysis (Section~\ref{sec:analysis}) includes a discussion on the impact of security advice followed at each lifecycle stage.
    
    \raggedbottom
    
    The Creation phase takes place under the authority of the manufacturer, where a device is designed, developed, and pre-configured. The Creation phase happens \textit{pre-deployment}, i.e., before the device is sold to an end-user. This excludes when a user receives the device from another user. 
    
    In the Installation phase, the user has received the device and readies it for normal use. This is the first post-deployment phase, and contains \textit{onboarding} or \textit{bootstrapping} (often used interchangeably or meaning slightly different things, depending on who uses it) \cite{IDBootstrap} which includes technical details such as key management, registration and identification of devices, establishing trust relationships, and other device configuration. 
    
    The Usage phase involves using the device as intended (e.g., a light bulb provides light, a smart thermostat controls temperature, a home security camera provides live camera access). While containing only two stages, the device is expected to spend most of its life in this phase. Software/firmware updates take place in this phase. 
    
    The Decommissioning phase is where a device ends its life with respect to a single user or organization. The device is readied for removal from its environment (data/key removal from the device), physically removed, and leaves the end-user's ownership (either via disposal or transfer of ownership to another user). If device ownership is transferred to another user, the device returns to the Installation phase where the post-deployment ownership phases begin again.
    
    \subsection{Established IoT Security Advice and the DCMS 1013 dataset}
    \label{sec:background/dcms}
    
    The IoT security advice analyzed in  Section~\ref{sec:analysis}, which we call the ``DCMS 1013-item dataset'' \cite{BellmanDataset}, is derived from an effort led by the UK government's Department for Digital, Culture, Media \& Sport (DCMS).  Their base dataset \cite{IoTSecMap} (Version 3 in particular ) consists of individual IoT security advice items compiled from earlier academic, industry, and government documents for manufacturers of IoT products \cite{DCMS1}. The dataset is positioned as security guidance rather than extremely detailed specification-level advice \cite{DCMS1}; we say more about this in Section~\ref{sec:bestpractices}. From a starting set of 1052 \textit{security advice items}, we manually removed duplicates---if two  items were word-for-word identical, we kept only one. This left 1013 items \cite{BellmanDataset}. The items originated from 69 documents (most were informal), from 49 different organizations. Groups represented in the collection include the IoT Security Foundation \cite{IoTSF1}, the European Union Agency for Network and Information Security (ENISA) \cite{ENISA1}, and the GSM Association (GSMA) \cite{GSMA4}.  While the most heavily referenced sources are industrial organizations whose main focus is security, others include the US Senate \cite{USSenate1}, the US National Telecommunications and Information Administration (NTIA) \cite{NTIA1}, and Microsoft \cite{Microsoft1}. As this suggests, a vast range of organizations have offered advice on IoT security. We chose this dataset for our analysis as it represents, to our knowledge, the most comprehensive publicly available collection of IoT security advice. 
    
    This 1013-item dataset is related to two other documents from DCMS, and one from the European Telecommunications Standards Institute (ETSI). For further context, we describe these here also. 
    
    \textbf{DCMS 13 guidelines document.} The DCMS \textit{Code of Practice for Consumer IoT Security} \cite{DCMS2} proposes 13 guidelines (these are also used in Australia \cite{AUSIoT}, with minor changes). Each includes a summary title and a more detailed guideline. The guideline descriptions typically do not express \textit{how} a guideline is to be executed (i.e., what specific steps to take), but rather an outcome or goal to reach. We note that this is contradicted by the guidelines also being positioned as ``practical steps'' \cite{DCMS2}, which leads us to expect something practical to do, i.e., an action to take rather than an outcome. We discuss \textit{actions} and \textit{outcomes} in Section~\ref{sec:background/actionvsoutcome}. 

    The DCMS 13 guidelines are intended to provide practical advice on securing IoT devices, and target four stakeholders: device manufacturers, IoT service providers, mobile application developers, and retailers \cite{DCMS2}. We note end-users are not mentioned as a targeted stakeholder. Table~\ref{tab:practices} lists the guideline titles along with our assignment of the IoT lifecycle stages in which they are applicable, and examples of (unrefereed) primary source documents suggesting such advice.

\begin{table}[t]
    \centering
    \caption{Assignment of DCMS guidelines to IoT device lifecycle phases. For full detailed guideline descriptions, see the DCMS 13 guidelines document \cite{DCMS2}.}
    \label{tab:practices}
    \footnotesize{
        \begin{tabular}{@{}p{1.0cm}p{5.0cm}p{1.2cm}p{1.5cm}@{}}
            \toprule
            & & Example  & Lifecycle\\
            %%%%%%%%%%%%%%%%%%%%%%%%%%%%%%%%%%%%%%%%%%%%%%%%%%%%%%%
            Guideline &
            Guideline Title &  
            Sources&
            Phase (Fig. \ref{fig:lifecycle}) \\
            \midrule
            %%%%%%%%%%%%%%%%%%%%%%%%%%%%%%%%%%%%%%%%%%%%%%%%%%%%%%%
            UK-1 &\textit{No default passwords} &
            \cite{CSA2, IEEE1, ENISA1} & 
            1.3 \\ \addlinespace[0.05cm]
            %%%%%%%%%%%%%%%%%%%%%%%%%%%%%%%%%%%%%%%%%%%%%%%%%%%%%%%
            UK-2 &\textit{Implement a vulnerability disclosure policy} &
            \cite{BITAG1,ENISA1,ACSCSS1} & 
            1.1 3.1a \\ \addlinespace[0.05cm]
            %%%%%%%%%%%%%%%%%%%%%%%%%%%%%%%%%%%%%%%%%%%%%%%%%%%%%%%
            UK-3 &\textit{Keep software updated}  &
            \cite{AIOTI1,BITAG1,CableLabs1} & 
            1.1 1.2b 3.1b \\ \addlinespace[0.05cm]
            %%%%%%%%%%%%%%%%%%%%%%%%%%%%%%%%%%%%%%%%%%%%%%%%%%%%%%%
            UK-4 &\textit{Securely store credentials and security-sensitive data} &
            \cite{CSA2,CSA1,ENISA1}  & 
            1.2a 1.2b 1.3 \\ \addlinespace[0.05cm]
            %%%%%%%%%%%%%%%%%%%%%%%%%%%%%%%%%%%%%%%%%%%%%%%%%%%%%%%
            UK-5 &\textit{Communicate securely}  &
            \cite{BITAG1,CSA1,ENISA1}   & 
            1.2b 1.3 \\ \addlinespace[0.05cm]
            %%%%%%%%%%%%%%%%%%%%%%%%%%%%%%%%%%%%%%%%%%%%%%%%%%%%%%%
            UK-6 &\textit{Minimise exposed attack surfaces}  &
            \cite{AIOTI3,ATT1,NYC2} & 
            1.2a 1.2b 1.3 \\ \addlinespace[0.05cm]
            %%%%%%%%%%%%%%%%%%%%%%%%%%%%%%%%%%%%%%%%%%%%%%%%%%%%%%%
            UK-7 &\textit{Ensure software integrity}  &
            \cite{ENISA1,GSMA4,IIC1}  & 
            1.2b \\ \addlinespace[0.05cm]
            %%%%%%%%%%%%%%%%%%%%%%%%%%%%%%%%%%%%%%%%%%%%%%%%%%%%%%%
            UK-8 &\textit{Ensure that personal data is protected} &
            \cite{AIOTI4,AIOTI3,NYC1} & 
            1.2a 1.2b 1.3 \\ \addlinespace[0.05cm]
            %%%%%%%%%%%%%%%%%%%%%%%%%%%%%%%%%%%%%%%%%%%%%%%%%%%%%%%
            UK-9 &\textit{Make systems resilient to outages}  &
            \cite{BITAG1,CableLabs1,ENISA1}   & 
            1.1 1.2b \\ \addlinespace[0.05cm]
            %%%%%%%%%%%%%%%%%%%%%%%%%%%%%%%%%%%%%%%%%%%%%%%%%%%%%%%
            UK-10 &\textit{Monitor system telemetry data}  &
            \cite{NYC2,CSA2,ENISA1}  & 
            1.1 1.2b 3.1a \\ \addlinespace[0.05cm]
            %%%%%%%%%%%%%%%%%%%%%%%%%%%%%%%%%%%%%%%%%%%%%%%%%%%%%%%
            UK-11 &\textit{Make it easy for consumers to delete personal data} &
            \cite{ENISA2,OTA1,IOTSF2}  & 
            1.1 1.2b \\ \addlinespace[0.05cm]
            %%%%%%%%%%%%%%%%%%%%%%%%%%%%%%%%%%%%%%%%%%%%%%%%%%%%%%%
            UK-12 &\textit{Make installation and maintenance of devices easy}  &
            \cite{ACSCSS1,IIC1,IOTSF2} & 
            1.1 1.2a 1.2b \\ \addlinespace[0.05cm]
            %%%%%%%%%%%%%%%%%%%%%%%%%%%%%%%%%%%%%%%%%%%%%%%%%%%%%%%
            UK-13 &\textit{Validate input data}  &
            \cite{ENISA1,OWASP2,IOTSF2}  & 
            1.2b \\ \addlinespace[0.05cm]
            %%%%%%%%%%%%%%%%%%%%%%%%%%%%%%%%%%%%%%%%%%%%%%%%%%%%%%%
            \bottomrule
        \end{tabular}
    }
\end{table}
    
    \textbf{DCMS mapping document.} The DCMS \textit{Mapping of IoT Security Recommendations, Guidance and Standards to the UK's Code of Practice for Consumer IoT Security} document \cite{DCMS1} maps each advice item in the 1013-item dataset (1052 items before our pre-processing, as noted above) to one of the 13 guidelines from the DCMS 13 guidelines document. While we do not analyze this document itself, we mention it for context, being referenced in both the DCMS 13 guidelines document and the ETSI provisions document (next). 
    
    \textbf{ETSI provisions document.} ETSI has also published a document of ``baseline requirements'' for IoT security \cite{ETSI2020} that appears to be an evolution of the DCMS 13 guidelines document. It includes all 13 categories as major headers (most with slightly modified wording), and elaborates on each with a set of requirements that fit the theme of that category, much as is done in the DCMS mapping document. For example, Provision 5.1-2 from the section titled \textit{5.1 No universal default passwords} (comparable to the DCMS' \textit{No default passwords} guideline header) states \cite{ETSI2020}:
    
    \begin{quote}
        \textit{Where pre-installed unique per device passwords are used, these shall be generated with a mechanism that reduces the risk of automated attacks against a class or type of device.}
        
        \vspace{4pt}
        \textit{EXAMPLE: Pre-installed passwords are sufficiently randomized.}
        
        \vspace{4pt}
        \textit{As a counter-example, passwords with incremental counters (such as ``password1'', ``password2'' and so on) are easily guessable. Further, using a password that is related in an obvious way to public information (sent over the air or within a network), such as MAC address or Wi-Fi SSID, can allow for password retrieval using automated means.}
    \end{quote}
    
    \noindent The majority of the advice items' topics therein are represented in the DCMS 1013-item dataset.
    
    While early portions of this paper deal primarily with terminology, these documents are introduced as examples of established IoT security advice, and in the case of the 1013-item dataset, we also analyze it in later sections.

\section{Defining what `Best Practice' Means}
\label{sec:bestpractices}

In this section we consider definitions for \textit{best practice}, including our own definition taking into account the concepts of outcomes, actions, and \textit{actionable} practices, and discuss related terms commonly appearing in the literature. Through this, we provide a refined, self-consistent vocabulary for security best practices and also disambiguate a wide variety of qualifying terms into three semantic categories.

    \subsection{Definition and Discussion}
    \label{sec:bestpractices/defn}
    The definition of best practice is largely taken for granted. Few documents that use it make any effort to explicitly define it. Of note, even RFC 1818/BCP 1 \cite{rfc1818}, the first of the IETF RFCs specifying what a Best Current Practice document is, fails to define best practice. Thus, the term (and concept of) best practice is, at least in security, almost always used casually, versus scientifically---the implicit assumption being that everyone understands what it means well enough to not require an explicit definition. 
    
    A negative consequence of this is that different experts also implicitly redefine \textit{best practice} to suit their own needs or context (examples are given in Section~\ref{sec:introduction} and the examples below supplement this). This leads to ambiguity, where certain uses of best practice have different meanings and connotations, while elsewhere different phrases may imply the same concept. To address this, we first propose a definition for \textit{practice} (separate from \textit{best} practice): \textit{A practice is a specific means intended to achieve a given desired outcome}. We build on this definition below. 
    
    A practice specifies actions (as explained in Section~\ref{sec:background/actionvsoutcome}) to reach an outcome, but does not necessarily imply any level of quality or security with respect to the means or mechanism used or the outcome (as discussed further in Section~\ref{sec:background/qualifying/quality}). Building on this definition, to reduce ambiguity and provide more precision to analyze security best practices, we propose the following practical definition for \textit{best practice}: \textit{For a given desired outcome, a best practice is a specific means intended to achieve that outcome, and that is considered to be better than, or at least as ``good'', as the best of other widely-considered means to achieve that same outcome}.\footnote{While one view of ``best'' might imply being above all known others, another is that ``best'' is a category that may have more than one member. It is thus reasonable to allow (by definition) that there are multiple best practices for a given desired outcome (consistent with King, above).}

    Note that by our definition, a best practice is something that can be done (an action), not something that is desired to be achieved (an outcome). A community in which a best practice is developed may have their own measure for quality, and quality requirements may vary based on the community, environment, and context. Further, as by our definition, best practices are specific to the outcome they aim to achieve, there is generally no ``silver-bullet'' best practice for use across all applications---practices typically must be tailored for the context \cite[Chapter 2]{Shostack2008} (e.g., a surgeon has different hand-washing best practices than an individual preparing a family meal). Best practices may be intended for manufacturers, but for the benefit of end-users (other stakeholders are often involved). Stakeholders that benefit may or may not be involved in a best practice's implementation. For example, a user of a product might not care how a manufacturer implements a best practice, despite relying on it for security.
    
    We now review four notable definitions of best practice and compare them to our definition above. 
    
    King \cite{King2000} provides extension discussion of security best practices, using the term \textit{best security practices} and giving particular attention to human aspects. In his view \cite{King2000}: 
    
    \begin{quote}
        {[Best practices are]} \textit{practices that have proven effective when used by one or more organizations and which, therefore, promise to be effective if adapted by other organizations.}
    \end{quote}
    
    \noindent King notes several central concepts, including that effectiveness is based on evidence of multiple instances (implying some degree of consensus and generality), that a practice must be applicable to real (not only theoretical) situations, and that it may exist among a set of others of equal quality for a given purpose \cite{King2000}.
    
    McGraw \cite{McGraw2006} discusses best practices in the context of development activities aiming to improve software security. He offers that best practices are: 
    
    \begin{quote}
        {[...]} \textit{usually described as those practices expounded by experts and adopted by practitioners.}
    \end{quote}
    
    \noindent In his view, best practices are created by a group of experts and often intended for use by non-experts. His book also regularly refers to \textit{touchpoints}, described as ``\textit{a set of software security best practices}'' \cite{McGraw2006}, implying that touchpoints \textit{are} best practices. However, discussion throughout the book indicates that these touchpoints are general categories of recommended processes and activities (e.g., code review or penetration testing \cite{McGraw2006}) rather than specific procedures that our definitions would recognize as actionable practices. 
    
    Garfinkel et al. \cite{Garfinkel2003} describe best practices (in the context of operating system and internet security) as: 
    
    \begin{quote}
        {[...]} \textit{a series of recommendations, procedures, and policies that are generally accepted within the community of security practitioners to give organizations a reasonable level of overall security and risk mitigation at a reasonable cost.}
    \end{quote}
    
    \noindent As a defining feature of a practice this emphasizes general agreement, within a community (however difficult that may be), on its quality. Note the hint that a best practice may involve a trade-off between cost and quality, some middleground acceptable for an organization. Related to this is \textit{feasibility}, discussed  in Section~\ref{sec:background/actionvsoutcome}. 
    
    If we consider the definitions by McGraw \cite{McGraw2006} and Garfinkel et al. \cite{Garfinkel2003} on a continuum of how specific they envision best practices to be (the scope of practices themselves), we find McGraw's description at one end (the coarse or general end), with Garfinkel et al.\ at the other end (specific). McGraw's best practices are comprised of general categories of activities to be executed at different stages of the software development lifecycle. Garfinkel discusses fine-grained practices, e.g., at the level of configuration commands, command-line arguments and precise details specific to a given OS and version (while not immediately clear from the quoted definition, this follows from examples throughout the cited book \cite{Garfinkel2003}.) 
    
    Shostack and Stewart \cite[pp.36--38]{Shostack2008}, in their book on security, describe best practices as:
    
    \begin{quote}
        {[...]} \textit{activities that are supposed to represent collective wisdom within a field} [and] \textit{designed to be vague enough to apply in the general case}.
    \end{quote}
    
    \noindent Our definition agrees regarding ``\textit{collective wisdom within a field}'', but we call for best practices to be more specific than vague. Shostack and Stewart's perspective fits somewhere between the earlier two on the above spectrum, being more specific than McGraw \cite{McGraw2006}, and less specific than Garfinkel et al. \cite{Garfinkel2003},
    
    While we avoid declaring these other characterizations of \textit{best practice} ``wrong'',\footnote{Absent a consensus definition for a term, how good or bad a particular definition is usually depends on how well it meets the purpose at hand.} herein we use our own definition to serve as a concrete, explicit reference, also allowing focus on concepts that we aim to highlight. We acknowledge that others may opt for a definition that does not in all cases require, as we do, a specific set of steps to reach a desired outcome. We require that beyond a goal (outcome) alone, a best practice includes a means to achieve a goal (preferring a specific set of steps). Our definition is more specific than informal views of best practices as lists of general ``good things to do''.
    
    One can also consider the implications of best practices from legal, technical, and social angles. From a legal perspective, following a best practice may be used as an argument to escape or limit liability, as in ``following the crowd'' or consensus as surely being reasonable. For example, financial institutions citing ``industry best practices'' to disclaim liability, per our example in Section~\ref{sec:introduction} \cite{CBC2020}. Technically, a best practice is often the best way known to technical experts or researchers for achieving an outcome (supported by some form of consensus), or as a way to limit risk \cite{Garfinkel2003}. Less formally, best practice often implies the most common (if not necessarily best) way to do something. At one level, one might argue that each of these are similar, but at a semantic level, they are different uses of the same term.
    
    \subsection{Outcomes vs. Actions}
    \label{sec:background/actionvsoutcome}
    \label{sec:background/actionable}
    
    We define an \textit{outcome} as the statement of the desired end goal that a stakeholder aims to reach, and an \textit{action} as an operation of one or more steps carried out by a person or computer, perhaps in order to achieve a desired outcome. For example, an outcome may be having created \textit{a strong password}, and an action to (partially) achieve this outcome may be to \textit{enforce a minimum password length of 8 characters} \cite{NISTsp80063}. In practice, outcomes or goals that are vague or broad may not give stakeholders a clear idea of any concrete set of actions that can be taken to achieve the goal. A desired outcome of ``strong security'', for example, is nebulous and cannot be mapped to specific actions to achieve the goal. Defining tightly-scoped outcomes or specifying an objective to withstand specific attacks allows for successful mapping to corresponding actions.
    
    A given practice may be viewed as \textit{actionable} if it can be carried out without guesswork by an advice target. We argue that being actionable is the crucial characteristic, the key point being to formulate advice such that the steps to be executed are explicit or well understood by targeted advice recipients. This leads to our next definition. By \textit{actionable practice} we mean a practice that involves a known, unambiguous sequence of steps, whose means of execution are understood (by the target advice recipients). 
    
    An explicit declaration (and characterization) of the target audience is also important, in our view. This impacts whether advice is actionable, as advice must often be tailored to an audience and their knowledge level. A sequence of steps described as \textit{generally understood} (in the above definition of \textit{actionable}) implies that the target audience has the appropriate level of knowledge to execute the advice. Advice not understandable by the target audience or not sufficiently specific becomes non-actionable to that audience. Wording and outcomes must be understood from both a (semantic) language and a technical perspective. For example, advice targeted at security experts (e.g., the AES specification \cite{FIPS197}) may be difficult to follow by non-expert audiences. Without a declaration of target audience, inappropriate audiences may attempt to
    execute advice and misinterpret or fail to successfully execute it, resulting in flawed implementations. 
    
    Advice that simply mentions a general technique by name (without details) is non-actionable, by our definition. However, pointers to ``next-level'' implementation details may meet our requirement of unambiguous steps (for actionability), e.g., with details in an external reference. In this way, advice may direct advice recipients to non-prescriptive techniques or approaches (e.g., key management techniques as in the UK-5 example below), but then link to further sources for specific details. This allows specifying how to carry out an advice item generally (how to approach a problem), while avoiding fine detail and lengthy descriptions, yet remaining actionable via links to detailed unambiguous steps. This avoids advice items that dictate an exhaustive number of individual steps, and avoids needing advice updates due to, e.g., parameter changes or algorithm upgrades; low-level details can be more frequently updated in external sources, without need to re-issue higher-level best practice advice (as it remains valid for longer periods).
    
    While we use \textit{actionable practice} (above) to emphasize that a practice must be one that a target subject can actually carry out, describing a practice as actionable is redundant, as all practices are necessarily actionable by our earlier definition of a practice as a \textit{specific means} to achieve a desired outcome. An outcome (alone) cannot be a best practice (or even a practice) as an outcome alone does not dictate a specific means to an end. In what follows, when we use the term \textit{practice}, we generally mean a practice that is actionable. 
    
    It follows that a recommendation specifying an outcome, but the path to which is an open research problem, cannot (and in our view should not) be considered a practice. Specifying ``advice'' that implies use of techniques that are experimental or unproven introduces ambiguity in how to carry out the advice and may result in inconsistent execution of the advice (which is not, by our definition, a practice). We argue that it is important for the security community---whether by academic, industrial, or government efforts---to identify and agree on practices with concrete desired outcomes for use by those targeted by the advice. Best practices adopted for specific use-cases will ideally lead to more reliable (correct) execution of the practices and thereby improve security. Our heavy focus on (actionable) practices arises from our belief that, if stated clearly, they may be the most promising, direct way to help pre-deployment stakeholders improve security.
    
    \label{sec:bestpractices/revisiting}
    Manually applying the above definition of \textit{actionable} to the full descriptions \cite{DCMS2} of each of the 13 DCMS guidelines whose titles are given in Table~\ref{tab:practices}, we found that only one of the guidelines fit our definition of actionable, that being \textit{UK-1}: ``\textit{No default passwords}''. This leads us to question how many of these guidelines can be reliably implemented from the guideline descriptions alone. For example, \textit{UK-5} (``\textit{Communicate securely}'') states \cite{DCMS2}:
    
    \begin{quote}
        \textit{Security-sensitive data, including any remote management and control, should be encrypted in transit, appropriate to the properties of the technology and usage. All keys should be managed securely. The use of open, peer-reviewed internet standards is strongly encouraged.}
    \end{quote}
    
    \noindent We find this guideline non-actionable, as it is vague and non-specific about which actions to take to follow it, and is unfocused on a single security topic (discussed in Section~\ref{sec:analysis/notuseful}). Guidelines may have implementation details inferred based on the experience of the implementer, but this does not appear to be the way these 13 guidelines are positioned. The associated DCMS mapping document \cite{DCMS1} is intended to provide additional details and context for how the guidelines should be followed, but as we later find (Section~\ref{sec:analysis}), the advice found in the mapping document is largely non-actionable.
    
    A practice does not necessarily need to specify a full sequence of low-level, specific, detailed steps; it may be acceptable to state high-level steps, provided they are still actionable. For example, a practice involving the use of AES does not necessarily require a line-by-line implementation as specified by NIST \cite{FIPS197}, but it may be enough for the practice to state a library or function to use, how it should be used, and specify any desired configuration details. To use a non-security example, a car mechanic does not need to build a car's alternator from scratch, but they are expected to be able to follow a guide to install and configure a pre-assembled one. This further highlights the importance of an appropriate target audience selection---depending on the context, specification-level details are important for, e.g., those building libraries and toolkits (to use the AES example), while others may require only the details needed to properly use the libraries provided to them. Both situations can have practices developed for them, while reaching desired outcomes and being appropriate for their respective audiences.
    
    While our definition of an actionable practice is designed to match what we expect is practically followable, clearly indicating what advice recipients must do, in some cases recipients can infer how to execute advice even if it lacks details. For example, depending on a target's experience, what a ``standard algorithm'' is may be understood. While we retain our definition of a practice being actionable, we acknowledge that in some cases, some advice recipients have sufficient experience to infer actionable detail from otherwise non-actionable advice---making explicit step-by-step instructions unnecessary. Nonetheless, because security experts are not always the audience  responsible for executing security advice (e.g., at an IoT device manufacturer), we encourage development of actionable practices for specific targeted audiences.
    
    \textbf{Infeasible advice.} We separate the concepts of a practice being actionable, and an implementer having the means by which to put said practice into place. Implementers must have the resources (technical, financial, personnel) available before a practice can be implemented, but availability of resources does not affect the generic actionability of a practice by our definition. (In other words: though a practice is actionable in general, that does not guarantee that a given party themselves has the resources to adopt the practice.) A practice that has a significant cost may be ruled out as a best practice by a recommending group, governing body, or peer community. Similarly, while still actionable by our definition, a practice that has (for example) 300 well-defined, unambiguous steps and takes 14 years to complete would likely not be considered as a best practice. Such a practice would be considered \textit{infeasible} (i.e., a practice that remains actionable, but viewed  as impractically inefficient by a party lacking the resources to carry it out). Note this is illustrated by the continuum of the actionability of practices (Fig.~\ref{fig:continuum-practices}, discussed later in Section~\ref{sec:methodology/flowchart}). An \textit{Infeasible Practice} (\textit{P3}) is actionable, but by fewer parties (suggested by its placement toward the \textit{actionable by fewer parties} labelled end of the Fig.~\ref{fig:continuum-practices} continuum) than a practice requiring a \textit{Security Expert} (\textit{P4})---as practically speaking, high costs reduce the number of parties able to implement a practice.
    
    \subsection{Imperative and Declarative Advice vs. Actions and Outcomes}
    \label{sec:bestpractices/declarative}
    
    By our definition (Section~\ref{sec:bestpractices/defn}), the statement of a best practice includes specifying a means to reach a desired outcome. We briefly consider now the utility of advice items that do not specify any such means or specific set of actions. As a particular case, consider an advice item that specifies an outcome whose attainment can be verified (but leaving it to an advice recipient to determine a specific means). Some advice recipients may still be able to attain the outcome, and auditors could verify attainment. Would such advice---which we call \textit{declarative} advice, next paragraph---be equivalent to a best practice? Not by our best practice definition, which requires a specific means; by our definition, an outcome and a best practice are categorically different. Nonetheless, if the means used to reach the outcome is of secondary importance to an advice giver or authority, and their primary interest is attaining the outcome, then advice items in the form of (verifiable) declarative outcomes may be useful alternatives to (actionable) best practices---for advice recipients who can independently determine a means to reach the outcome. Having made this observation,\footnote{
        We thank an anonymous referee for raising this question.
    } we proceed herein to use our (Section~\ref{sec:bestpractices/defn}) definition of best practice.
    
    As supporting context, we note that actions and outcomes can respectively be mapped to \textit{imperative advice} (advice that includes specific steps or actions to reach an outcome), and \textit{declarative advice} (advice that specifies an end result or outcome to reach, but not any specific method by which to reach it) \cite{Boley1991, Fahland2009}. Depending on their nature, some outcomes may be verifiable, e.g., through a test that yields a yes/no answer to whether the goal was reached, or a measure used against a pass-fail threshold. For example, consider the advice item \cite{IOTSF2}: \textit{Where a device or devices are capable of having their ownership transferred to a different owner, all the previous owner's Personal Information shall be removed from the device(s) and registered services}. This could be verified, for example, by checking that any memory region designated for storing user personal information has zeros in every byte.
    
    In contrast, an example of a non-verifiable advice item is \textit{use a randomly generated salt with a minimum length of 32 bits for hashing with passwords} \cite{NISTsp80063}. If we assume a verifier is only presented with the fixed-length output from a hashing function (i.e., $h$ from $h = H(p,s)$, where $p$ is a password and $s$ is a salt value), this practice is not verifiable, as the output provides no indication of the salt's length or method of generation. For example, a password of ``$password123$'' and a salt of ``$ABCD$'' produces a SHA3-256 hash of ``$e0aa...beef$'' (truncated), which does not reveal any characteristics of the salt. 
    While we use this as an example, in practice, salts are typically stored with the hash output---if a verifier were to recover one, they would then have access to the other.

\newcommand{\LenClass}{1.2cm}
\newcommand{\LenQual}{0.6cm}
\newcommand{\LenTerm}{3.4cm}

\newlength\LenMax
\setlength\LenMax\textwidth
\addtolength\LenMax{-\LenClass}
\addtolength\LenMax{-\LenQual}
\addtolength\LenMax{-\LenTerm}
\addtolength\LenMax{-1.2cm}

\setlength{\tabcolsep}{5pt}
\renewcommand{\arraystretch}{1}

\begin{table}[t]
    \centering
    \caption{Categories of commonly used qualifying terms related to best practices.}
    \label{tab:qualifyingterms}
    \footnotesize{
        \begin{tabular}{@{}p{\LenClass}p{\LenQual}p{\LenTerm}p{\LenMax}@{}}
            \toprule
            Category Focus & 
            &
            Qualifying Terms (examples) & 
            Suggested Use \\ \midrule
            %%%%%%%%%%%%%%%%%%%%%%%%%%%%%%%%%%%%%%%%%%%%%%%%%%%%%%%
            \multirow{1}{\LenClass}{Quality} & \multirow{2}{\LenClass}{\"{U}ber} & ``state-of-the-art'' & \multirow{2}{\LenMax}{For practices considered superior to all others, even if not widely adopted. These terms imply elite quality, possibly at high cost or complexity.}\\
            & &``gold standard'' & \\
            \addlinespace[0.05cm] \cline{2-4} \addlinespace[0.05cm] 
            %%%%%%%%%%%%%%%%%%%%%%%%%%%%%%%%%%%%%%%%%%%%%%%%%%%%%%%
            & \multirow{2}{\LenClass}{Best} & ``best current practice'' & \multirow{1}{\LenMax}{For practices \textit{widely-considered} to be high quality (plus widely adopted, ideally).}\\
            & &``best practice'' & \\
            \addlinespace[0.05cm] \cline{2-4} \addlinespace[0.05cm] 
            %%%%%%%%%%%%%%%%%%%%%%%%%%%%%%%%%%%%%%%%%%%%%%%%%%%%%%%
             & \multirow{3}{\LenClass}{Good} & ``recommended practice'' & \multirow{3}{\LenMax}{For practices that are beneficial (e.g., to improve security), without implying that better practices do not exist. Here, ``recommended'' and ``suggested'' do not imply a formal endorsement.}\\
            & &``suggested practice'' & \\ 
            & &``good practice'' & \\ 
            \addlinespace[0.05cm] \midrule
            %%%%%%%%%%%%%%%%%%%%%%%%%%%%%%%%%%%%%%%%%%%%%%%%%%%%%%%
    		Commonality & &``minimum expectation''& \multirow{3}{\LenMax}{For practices not necessarily implying quality, but reflecting wide use. Alternatively, these may be de facto practices or functionality, informally recognized by experts as generally expected.}\\
    		& &``baseline practice'' & \\
    		& &``accepted practice'' & \\
    		& &``common practice'' & \\
    		& &``standard practice'' & \\
    		\addlinespace[0.05cm] \midrule
    		%%%%%%%%%%%%%%%%%%%%%%%%%%%%%%%%%%%%%%%%%%%%%%%%%%%%%%%
    		Stipulation & &``regulation'' & \multirow{3}{\LenMax}{For practices endorsed (formally) or mandated in some capacity by an organization or individual. Includes practices that may be, in some way, enforced by an entity such that there implies a negative consequence if the advice is not followed.}\\
    		& &``mandatory practice/requirement'' & \\ 
    		& &``formal standard'' & \\
    		& &``code of practice'' & \\
    		& &``recommendation'' & \\ 
    		& &``guideline'' & \\ 
    		\addlinespace[0.05cm]
    		%%%%%%%%%%%%%%%%%%%%%%%%%%%%%%%%%%%%%%%%%%%%%%%%%%%%%%%
            \bottomrule
        \end{tabular}
    }
\end{table}
    
    \subsection{Commonly-Used Qualifying Terms}
    A number of what we call \textit{qualifying terms} are widely used as an adjective before the word practice (e.g., \textit{common}, \textit{good}, \textit{best}) but without definition of the qualifying term itself. Being widely used might suggest that readers know (and are in universal agreement on) what authors mean when they use these terms. Like \textit{best practice}, while security community members are apparently expected to have a general intuitive understanding of the meanings of these terms, consensus has not been reached on the meanings of these terms either. 
    
    For example, IETF BCP draft \textit{Best Current Practices for Securing Internet of Things Devices} \cite{Moore2017} contains advice that is arguably positioned in three different ways: (1) as advice within a \textit{best current practices} document (containing advice considered to be the best current practices); (2) as \textit{recommendations} (suggesting that use of the advice items is endorsed); and (3) as \textit{minimum requirements} (their use is a minimum expectation).
    
    In an effort to both highlight existing terminology and move toward more consistent use of terminology, we associate these highlighted terms (among others) with one of three distinct categories of qualifying terms, summarized in Table~\ref{tab:qualifyingterms}: \textit{quality}, \textit{commonality}, and \textit{stipulation}. These three categories, discussed in separate subsections below, can be used to characterize a given advice item. Table~\ref{tab:qualifyingterms} also suggests where/when each qualifying term should be used and gives examples. Greater consistency in use of terminology may reduce ambiguity, misunderstandings, and consequent errors, within the academic and industrial security communities. From an advice recipient's perspective, it may also clarify the expected outcomes of following advice (e.g., what outcomes will be reached), and expectations surrounding its use (e.g., how the advice should be carried out). 
    
    While we primarily categorize terms by what we view as each term's dominant goal (i.e., identifying the quality of an advice item, how commonly an advice item is used, and acknowledging a governing authority's stipulation of the advice item), an advice item can share the characteristics of more than one category. For example, an advice item that is considered to be a \textit{good practice} (quality category) can also be a \textit{standard practice} (commonality category) through wide use, and a \textit{best practice} (quality) can be included in a formal standard (stipulation). 
    
    Table~\ref{tab:qualifyingterms} does not explicitly define the commonly-used qualifying terms contained therein; rather, it describes how we suggest each term (belonging to a category) be most appropriately used. For example, here our Suggested Use for \textit{best practice} expresses its relationship to being widely considered of high quality (albeit a higher quality tier exists), while our definition (Section~\ref{sec:bestpractices/defn}) explicitly notes that best practices are better than (or at least as good as) other high quality practices with wide consideration. In what follows, we discuss these terms in greater detail.  
    
    \subsection{Category 1: Quality-based Terms}
    \label{sec:background/qualifying/quality}
    Quality-based terms provide a natural basis on which to differentiate practices. Conceptually, we order \textit{\"{u}ber}, \textit{best}, and \textit{good} practices along a quality continuum. We note that terms used to describe practices of low quality (i.e., below good) receive less attention in literature as documents promoting security advice focus more on good than bad practices. Our definition of a good practice (the lowest quality we formally recognize) implies that anything lower does not improve security. 
    
        \textbf{\"{U}ber practices.} 
        The sub-category or group \textit{\"{U}ber} suggests practices that are in some way superior to best practices, or beyond what would be considered already high quality. \textit{State-of-the-art} or \textit{gold standard} implies something of elite technical quality, but perhaps not yet widely adopted. Consider as a practical example: in luxury cars, a heated steering wheel. While more comfortable on a cold winter day, best practice would likely be to ensure correct function and adequate steering grip to reduce the likelihood of accidents. A heating function may be the ``gold standard'' or ``state-of-the-art'' (typically at higher cost).
        
        \textbf{Best practices.} 
        The group \textit{Best} suggests practices widely considered to be high quality, and often, widely adopted. While technically better practices may exist, best practices are widely accepted within a community to be high quality. 
        
        \textbf{Good practices.} 
        The group \textit{Good} suggests practices that improve security but are not necessarily the best practices available. They generally are not lauded for high quality per se. A good practice often either does not have wide acceptance as being the best, or is perhaps not widely practiced or not considered essential even if easy and beneficial (e.g., a good practice is to apply the emergency brake when parking facing down a hill, while a best practice is to both apply the emergency brake \textit{and} turn the wheels to the curb). Further context may prove useful for understanding their use. For example, access control to a low-value free online newspaper account may not require a best practice authentication method (per our definition); a good practice may suffice \cite{Garfinkel2003}. In other words \cite{King2000}: ``\textit{sometimes the good is good enough}''.
        
    \subsection{Category 2: Commonality-based Terms}
    \label{sec:background/qualifying/commonality}
    Commonality-based terms also often include the word \textit{practice} (e.g., \textit{accepted practice}, \textit{common practice}), but their unifying trait is frequency of use rather than quality.
    
    \textbf{Baseline practice/minimum expectation.} 
    These terms suggest a minimum level to be reached. We assign these to the commonality category, as it is expected that the minimum acceptable level of advice is commonly followed.

    \textbf{Common/standard/accepted practices.} 
    These terms reflect broad usage. For example, it may (unfortunately) be common to store passwords in plaintext within a database (thus being a common practice), but that is not best practice (or even a good practice).
        
    We repeat that commonality does not necessarily imply quality. Terms in this category are less clearly ordered than in the quality category, and some terms are used interchangeably (e.g., \textit{common}/\textit{standard}/\textit{accepted}). As \textit{baseline} and \textit{minimum expectation} both imply a lowest reasonable threshold to start from, these may be considered more of a priority to be followed, thus we order them higher in the group than the \textit{common}/\textit{standard}/\textit{accepted} practices. 
    
    Correlated with commonality is the \textit{maturity} of advice, typically reflecting the length of time that advice has been, or continues to be, given or known. To follow an earlier example, while not considered even a good practice, storing passwords in plaintext has become a mature practice \cite{PlainTextOffenders}. 
    Ideally, a best practice would be mature as well as widely considered to be high quality (Section~\ref{sec:background/qualifying/quality}), but greater maturity of an advice item does not always imply higher quality (e.g., DES is a mature cipher, but no longer best practice). 
    
    \textbf{Security design principles.} Security design principles are a known set of guiding rules which aim to improve security \cite{Saltzer1975}. These principles are generally based on experience, suggesting their maturity. Security design principles are also generally expected (by experts) to be followed, and are complementary to the existing categories, but we intentionally omit them from Table~\ref{tab:qualifyingterms}, and discuss them further in Section~\ref{sec:methodology/principles}.
    
    \subsection{Category 3: Stipulation-based Terms}
    \label{sec:background/qualifying/stipulation}
    Distinct from quality and commonality, some terms related to best practices have more to do with the endorsement by an authority, the authority's jurisdiction, and whether the advice is mandatory (i.e., a firm requirement). Note that the entity creating advice is not necessarily the authority mandating its use. Our \textit{stipulation} category contains qualifying terms describing advice that is mandated or endorsed by an entity in some way. These too can be ordered along a continuum. On the strict end are terms that imply a negative consequence for not following the advice (e.g., \textit{mandatory practice}, \textit{requirement}, \textit{regulation}). On the looser end are terms that are stipulated, but not necessarily enforced (e.g., \textit{guideline/guidance}, \textit{recommendation}). As with a best practice, stipulations should, in our view, ideally be accompanied by an explanation of the intended outcome.
    
        \textbf{Regulation.} We use \textit{regulation} to mean a directive from an authority stating specific advice that must be followed to be allowed to operate within a \textit{jurisdiction}. Here, a jurisdiction refers to the legal or authoritative domain, or the context of the deployment environment or use cases (e.g., home IoT may require different practices than IoT devices for government; physical locations, e.g., to meet certain requirements to be allowed to be sold in a country; or scope of technology, e.g., certain practices may be more appropriate for IoT devices rather than desktop computers). 
        
        \textbf{Mandatory practices/requirements.} Hereafter just ``\textit{requirements}'', these do not necessarily imply the quality of a given practice, but rather that it is stipulated by some governing body or regulation, suggesting official endorsement. These may be considered ``enough'' for some purposes (e.g., \textit{enough to not be sued} or \textit{enough to pass inspection}). Practices across a range of qualities may be requirements depending on the governing body or motivation, although a practice established as high quality is more likely to become a requirement.
        
        \textbf{Formal standard.} 
        We take \textit{formal standard} to mean a formally documented (endorsed by some authority) specification. This typically (but not necessarily) implies acceptable quality; the main point is to officially specify details and recognize, e.g., a particular method or measurement. The purpose of a standard may be interoperability---e.g., standards for the gauge of rail tracks or pipes. In this context, formal standard differs from \textit{standard practice} (i.e., common practice, above) and is not related to frequency of use, e.g., it is \textit{standard} (practice) to eat at 12 noon. Standards are typically sufficiently detailed such that conformance or compliance can be judged by, e.g., an auditor, or interoperability tests \cite{Garfinkel2003}.
       
        \textbf{Code of practice.}
        We take \textit{code of practice} to mean a set of guidelines designed to help inform others (traditionally within a profession) of expectations. They often pertain to ethical or safety issues. Codes of practice are often stipulated (within an organization or industry), but may be viewed as voluntary in that, e.g., failure to follow them typically does not result in major penalties unlike stricter terms (mandatory practices/requirements). In our use, a code of practice is distinct from formal regulations such as an ``electrical code'' or ``building code''.

        \textbf{Recommendations and guidelines.} 
        A \textit{recommendation} is an endorsement of, e.g., a practice by an individual or organization as their suggested way to do something. Recommendations (depending on the recommending entity) may be subject to bias or be self-serving, and do not necessarily reflect expertise or universal consensus. Some recommendations, depending on their sources, are, in essence, requirements. Recommendations commonly suggest following a standard \cite{DCMS1}. Similarly, a \textit{guideline} or \textit{guidance} is often given to promote a suggested way to achieve a goal (or as described by Garfinkel et al. \cite{Garfinkel2003}, something that \textit{should} be done). A guideline may be used in the spirit of a recommendation---offered as help, versus imposing rules.

As final thoughts on best practices and related terminology, we noted inconsistency in the use of common terms (Section~\ref{sec:bestpractices}) and the situations in which they are used, with the same terms used with different meanings in different situations. We argue that supporting consistent use of terms, and separating the concepts of quality, commonality, and stipulation provides a better foundation to discuss and analyze security advice within the community. Additionally, we discussed what we argue is an important characteristic for security advice: whether it is actionable.

\section{Security Advice Coding Tree Methodology and Development}
\label{sec:methodology}

Our methodology for the systematic analysis of existing IoT security advice begins with iterative inductive coding. Our specific goal is to categorize (\textit{code}) 1013 IoT security advice items\footnote{
    This is not to be confused with the DCMS 13 guidelines detailed in Section~\ref{sec:background/dcms}.
} for further analysis and thereby to characterize the current state of IoT security advice. Such categorization is commonly done on, e.g., qualitative data from interview responses, to allow further analysis. The iterative inductive coding process begins with reviewing and understanding the dataset and developing codes (collectively, a \textit{codebook}) to assign to dataset items, and iteratively refining the codes as new themes, relationships, and insights emerge from further review \cite{Thomas2006}. Inductive coding techniques are commonly used in computer science and security research in the analysis of qualitative data (e.g., \cite{Krombholz2017, Kang2019, Naiakshina2017}). 

    \subsection{Establishing Analysis Tools}
    \label{sec:methodology/flowchart}
    
    \textbf{Establishing an initial codebook.}  
    To begin development of a codebook for inductive coding, a first \textit{coder} (C1)\footnote{\label{foot:coders}
        A \textit{coder} is a researcher that conducts iterative inductive coding as described in this section. The three coders (C1--C3) described are the three authors. 
    } initially reviewed the 1013-item IoT security advice dataset (Section~\ref{sec:background/dcms}) to extract unrefined categories (\textit{codes}) that characterize advice items. Discussion and preliminary test codings based on the extracted categories by the first and second coder (C2) resulted in the following coarse codebook (set of codes): \textit{Practice}, \textit{Incompletely specified practice}, \textit{Outcome}, \textit{Security design principle}, \textit{Too vague to tell}, \textit{Out of scope}.
    
    After establishing this early codebook (with associated definitions; the final codebook definitions are discussed in the next paragraph), test sets consisting of ten new, mechanically selected items from the dataset (to test across topics within the dataset; e.g., item numbers 100, 200, [...], 1000) were coded to determine agreement between the two coders. This consisted of each coder reading, interpreting, and assigning each item in the test set to a code (informally called a \textit{tag}). This process was done a second time on a distinct test set of 10 items. From this process we refined the codebook by creating new codes for items that did not fit well into existing codes. Assigning an item directly to one of the 6 codes resulted in low inter-coder agreement of between 30--40\% for the first two test sets. This motivated the development of the coding tree (described next).
    
    \textbf{Establishing the coding tree.} 
    In an effort to reduce subjectivity, what we call a \textit{coding tree} was built to more objectively guide coders toward codes based on a sequence of \textit{yes}/\textit{no} questions (the final version of the coding tree is Fig.~\ref{fig:flowchart2}; the final codebook is in Fig.~\ref{box:tags}).\footnote{
        To our knowledge, we are the first to build a coding \textit{tree} that uses questions to guide coders to tags for qualitative data. Typically, a codebook is built iteratively and then used directly (e.g., \cite{Huaman2021, Krombholz2017, Kang2019, Naiakshina2017}).
    } For a given advice item, starting at the top of the tree, each question progressively directs coders to a next question via branches down the tree, finally arriving at a leaf node (containing the resulting code). An additional coder (C3) was used to assist in test codings and further refinement of the coding tree. 
    
    This method was iteratively refined through five test coding sets, where coders refined the codebook codes, questions, and organization of the tree. Improvement was measured based on inter-coder agreement after modification (i.e., addition/modification of codes and/or questions) of the tree. Over several iterations, coders discussed the results to resolve ambiguities and gaps in code definitions or questions in the coding tree, refining questions difficult to reliably answer and coming to an agreement on the codes \cite{Huaman2021}, improving inter-coder agreement through a relatively concise decision path. To code items that required more reflection, coders consulted a further detailed annotation (see App.~\ref{app:cheatsheet}), which was created by the first coder during the refinement phase and used by coders during the full coding exercise. 
    
\begin{figure}[h]
        \centering
        \includegraphics[width=0.57\textwidth]{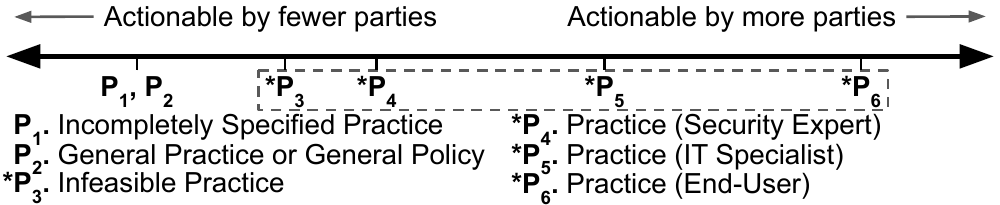}
        \caption{Actionability continuum: practice categories. Terms defined in Fig.~\ref{box:tags}. An asterisk (*) indicates categories we define as actionable.}
        \label{fig:continuum-practices}
    \end{figure}

    Many advice items positioned as practices in the DCMS 1013-item dataset were explicit about technique or technical method to address security, but lacked actionable detail. For example, Item \#50 in the 1013-item dataset \cite{IIC1} states: \textit{Endpoints must always use standard cryptographic algorithms}. This led us to develop the continuum of Fig.~\ref{fig:continuum-practices}. On its left side, practices are less widely actionable (an incompletely specified practice, a general practice/policy). These often specify vague technical directions to take or methods to use (``\textit{standard cryptographic algorithms}'' in the example), but not explicit actions as typically needed to allow successful execution. On the right side are practices that even end-users would be able to carry out (\textit{P6}), suggesting that if an end-user would be able to carry out the practice, so would a more experienced implementer. Moving left requires more in-depth knowledge and experience to understand (or infer a direction from) a practice, and implementation details become more ambiguous or unclear (even to a security expert). Coders did not use this continuum for coding, but we use it to visually represent where each category of practice may exist in relation to each other as a companion to the coding tree.
    
    \raggedbottom
     
    The binary \textit{yes}/\textit{no} decisions made by coders (through use of the coding tree) resulted in codes being assigned to advice items; as a side effect, by reaching one of the codes also in Fig.~\ref{fig:continuum-practices}, advice items were indirectly placed onto the continuum. In contrast, where to directly (manually) place an advice item on the continuum (\textit{P1}--\textit{P6} in Fig.~\ref{fig:continuum-practices}) may be less clear or might result in some point between two codes. For example, Item \#90 states \cite{IOTSF2}: \textit{Communications protocols should be latest versions with no publicly known vulnerabilities and/or appropriate for the product}. This advice item might be (manually, i.e., directly) coded as either \textit{P4} or \textit{P5} depending on knowledge of the coder. 
    
    \textbf{Making a second code available to coders.} 
    To further address reproducibility in use of the coding tree, a second code (for a given advice item) is optionally available to a coder. If the coder reaches a question that, based on their interpretation of the item, could be answered both as \textit{yes} and \textit{no}, this option allows both the \textit{yes} and \textit{no} edges in the coding tree to be followed to their respective leaf node code.\footnote{
        Coders were limited to at most two codes per advice item (i.e., at most a single extra \textit{yes}/\textit{no} question answer was allowed for any one advice item).
    } As a result, both codes (\textit{first} and \textit{second} codes) would be assigned to the advice item. For example, Item \#977 states \cite{PSACertified1}: \textit{The RTOS makes use of secure storage to protect sensitive application data and secrets and additionally binds the data to a specific device instance.} A coder may answer \textit{no} to Question 5 if they believe this advice does not describe or imply actions to take (thus resulting in a code of \textit{Incompletely Specified Practice}, \textit{P1}), but if they could argue it does, they could answer \textit{yes} to Question 5, leading them toward codes \textit{P3}--\textit{P6}. As an argument could be made by a coder that a question could be answered \textit{yes} \textit{and} \textit{no}, we made the decision to consider both codes as equals in our analyses, i.e., neither code is considered more or less important than the other; both codes assigned to an advice item are counted in the results (Section~\ref{sec:analysis}). For calculating agreement in trial codings, in cases where two coders availed themselves to a second coding on the same item, if coders agreed on at least one code, it was counted as an agreement for that item.
    
    \begin{figure}[b]
        \centering
        \includegraphics[width=0.57\textwidth]{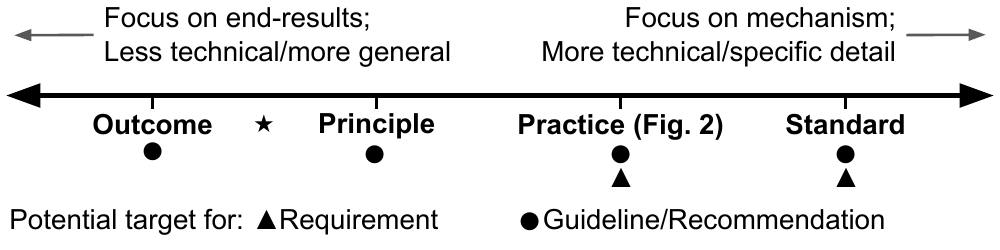}
        \caption{Relationship between terms based on inferred focus of advice's intent. The $\star$ denotes where we suggest policies \cite{Garfinkel2003b} fit on this continuum; arguably, they might alternatively be located in parallel or just to the right of principles.}
        \label{fig:continuum}
    \end{figure}
    \raggedbottom
    
    \begin{figure}[ht]
        \noindent
        \fbox{\parbox[t]{0.98\textwidth}{
            \small{
                \vspace{2pt}
                \textbf{~P1. Incompletely Specified Practice:} Advice that suggests a technical direction of a practice (e.g., a technical method/technique, software tool, specific rule), but lacks clear indication of any steps to be taken, and fails to meet our definition of actionable.
                \vspace{2pt}
                
                \textbf{~P2. General Practice or General Policy:} Advice that is not explicit about any techniques or tools, but is considered a general approach to improving security. This may also be policy-related advice. These are not considered actionable (despite being labeled as a practice) due to their general, less specific nature. 
                \vspace{2pt}
                
                \textbf{*P3. Infeasible Practice:} A practice, but one whose execution would require an unreasonable amount of resources (e.g., time, financial, human), or cost vastly more than what benefit would be gained.
                \vspace{2pt}
                
                \textbf{*P4. Specific Practice---Security Expert:} A practice requiring an expert in security to execute. These may require in-depth knowledge and experience of security topics, and often rely on the advice recipient to infer steps that are not clearly defined in the advice.
                \vspace{2pt}
                
                \textbf{*P5. Specific Practice---IT Specialist:} A practice that IT specialists (dedicated IT and development employees) developing or maintaining a product would be able to execute. These practices do not require the advice recipient to be a \textit{security} expert, but assumes basic knowledge of computer security such as that gained through coursework in formal or informal education.
                \vspace{2pt}
                
                \textbf{*P6. Specific Practice---End-User:} A practice an end-user would be able to execute. These are actionable by that audience, and typically executed by the user via direct interaction with the device, using a mobile app, or cloud service.
                \vspace{2pt}
                
                \textbf{~N1. Security Principle:} Advice that suggests a generic (as in applying to many situations) rule to follow that has shown through experience to improve security outcomes or reduce security exposures.Discussed in Section~\ref{sec:methodology/principles}. 
                \vspace{2pt}
                
                \textbf{~N2. Security Design Principle:} Advice that suggests a \textit{Security Principle}, but specifically for the Design phase of the lifecycle (therefore a subset of \textit{security principle}).
                \vspace{2pt}
                
                \textbf{~O1a/b. Desired Outcome:} Advice that suggests a generic, high-level end goal that a stakeholder would like to attain (as opposed to a means by which to reach a goal).
                \vspace{2pt}
                
                \textbf{~M1. Not Useful (too vague/unclear or multiple items):} Advice that does not make sense from a language perspective (e.g., not full sentence, unclear grammar), or is not focused on a specific task/action to complete.
                \vspace{2pt}
                
                \textbf{~M2. Beyond Scope of Security:} Advice that is not clearly an item that would be implemented for the benefit of security.
            }
        }}
        \caption{Codes and descriptions for coding tree of Fig.~\ref{fig:flowchart2}. As discussed (Section~\ref{sec:analysis/notuseful}), coders who reach \textit{M1} for an item may use an optional sub-label to denote the item as \textit{Unfocused}. An asterisk (*) indicates categories we define as actionable (matching Fig.~\ref{fig:continuum-practices}).}
        \label{box:tags}
    \end{figure}
    \raggedbottom
    
    \begin{figure}
        \centering
        \subfloat{{\includegraphics[width=0.55\textwidth]{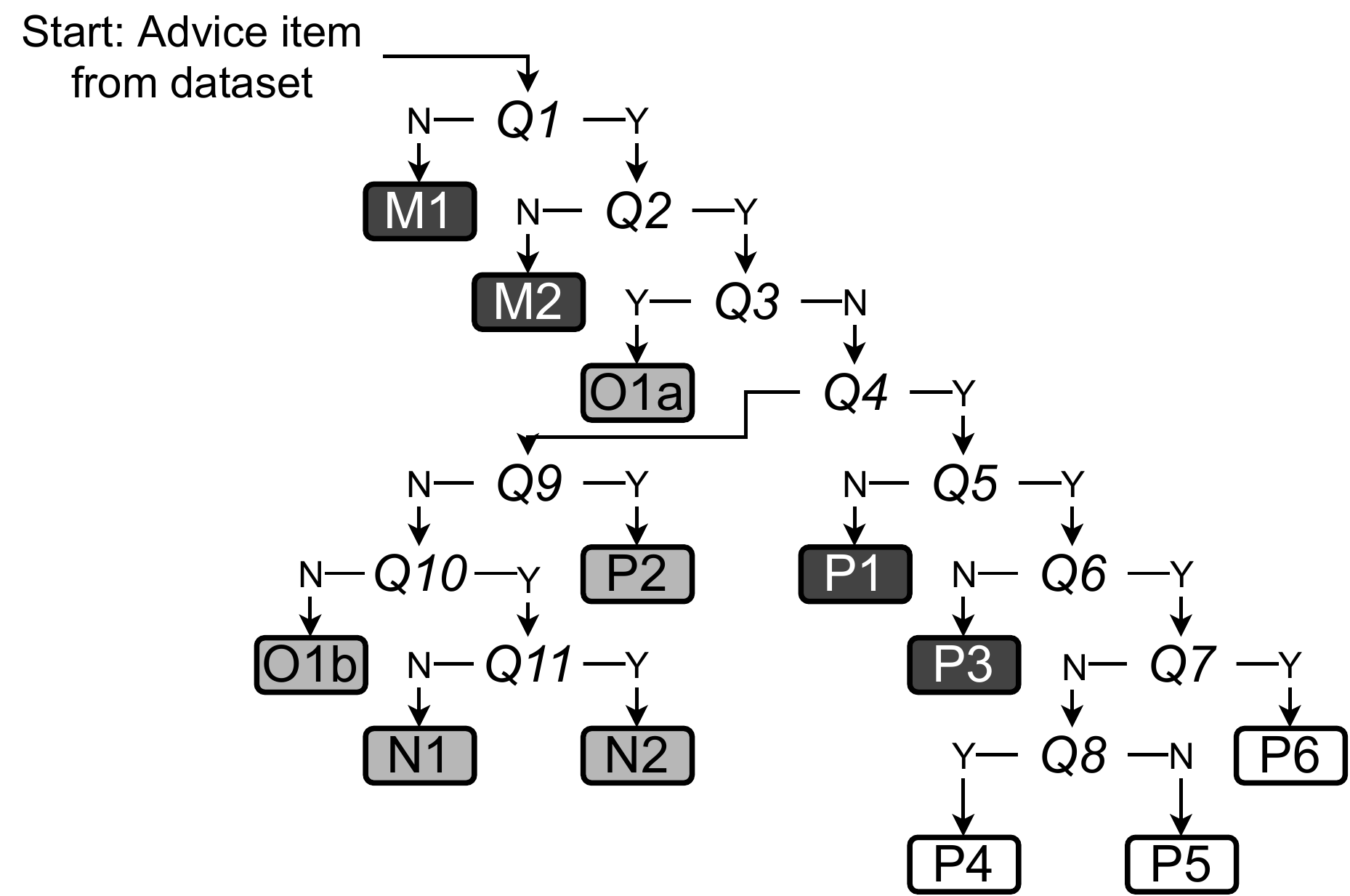}}}
        \qquad
        \subfloat{{
            \small{
                \parbox[t]{0.7\textwidth}{
                    \newcommand{\Spacing}{-3pt}
                    \vspace{\Spacing}
                    \paragraph{Q1.} Is the item conveyed in unambiguous language, and relatively focused?
                    \vspace{\Spacing}
                    \paragraph{Q2.} Is it arguably helpful for security?
                    \vspace{\Spacing}
                    \paragraph{Q3.} Is it focused more on a desired outcome than how to achieve it?
                    \vspace{\Spacing}
                    \paragraph{Q4.} Does it suggest a security technique, mechanism, software tool, or specific rule?
                    \vspace{\Spacing}
                    \paragraph{Q5.} Does it describe or imply steps or explicit actions to take?
                    \vspace{\Spacing}
                    \paragraph{Q6.} Is it viable to accomplish with reasonable resources?
                    \vspace{\Spacing}
                    \paragraph{Q7.} Is it intended that the end-user carry out this out?
                    \vspace{\Spacing}
                    \paragraph{Q8.} Is it intended that a security expert carry out this item?
                    \vspace{\Spacing}
                    \paragraph{Q9.} Is it a general policy, general practice, or general procedure?
                    \vspace{\Spacing}
                    \paragraph{Q10.} Is it a broad approach or security property?
                    \vspace{\Spacing}
                    \paragraph{Q11.} Does it relate to a principle in the design stage?
                }
            }
        
        }}
        \caption{Decision tree for assigning codes to advice items (coding tree). Leaf node codes explained in Fig.~\ref{box:tags}. Black shading (\textit{M1}, \textit{M2}, \textit{P1}, \textit{P3}) denotes advice considered not helpful to include (for lack of actionability or feasibility); white codes (\textit{P4}, \textit{P5}, \textit{P6}) are desirable actionable practices (excluding infeasible \textit{P3}); grey-codes (\textit{O1a}, \textit{O1b}, \textit{N1}, \textit{N2}, \textit{P2}) are considered useful context, but non-actionable.
        }
        \label{fig:flowchart2}
    \end{figure}
    
    \textbf{Determining test set inter-coder agreement.} 
    For the development of the coding tree, i.e., during test codings, as opposed to the full coding of the 1013-item dataset, by convention, we considered agreement between two test coders if:
    
    \begin{itemize}
        \item their coding resulted in agreement on at least one code (one coder's \textit{first} or \textit{second} code matches either the \textit{first} or \textit{second} code of the other); OR
        
        \item both coders' decisions resulted in an item coded into any code category from \textit{P1} to \textit{P6} per Fig.~\ref{fig:continuum-practices}; AND those two codes were within plus-or-minus one code distance away on the Fig.~\ref{fig:continuum-practices} continuum.
    \end{itemize}
    
    \noindent For example, if one coding was \textit{Infeasible Practice} (\textit{P3}) and the other \textit{Specific Practice---Security Expert} (\textit{P4}, one position right of \textit{P3}), we declared this an agreement on the basis that their proximity on the continuum implies equivalence, taking into account the subjective nature of coders answering the decision questions. We refer to this as the ``plus-or-minus one rule'' (``$\pm$1''). For advice items where at least one coder used a second code (resulting in a total of 3 or 4 codes on an advice item---2 from one coder and 1 from the other, or 2 from both coders), we counted it as an agreement if either code from a coder was within $\pm$1 distance from one of the other coder's codes. 
    Fig.~\ref{fig:continuum} relates practices in our continuum (Fig.~\ref{fig:continuum-practices}) with other concepts.\footnote{
        This is discussed further in Section~\ref{sec:methodology/principles}.
    }
    
    \textbf{Final test set coding and inter-coder agreement.} 
    A final test coding was done with a set of 20 items. Based on first/second codes and $\pm$1 rule, the mean agreement rate between the three coders was 73\% (C1 and C2, 80\% agreement, Cohen's Kappa \cite{McDonald2019} $\kappa = 0.74$; C1 and C3, 75\%, $\kappa = 0.67$; C2 and C3, 65\%, $\kappa = 0.59$). After the final test coding by the three coders, a detailed technical analysis and full coding of the 1013 item dataset was done by the first coder using the coding tree of Fig.~\ref{fig:flowchart2}. 
    
    In contrast to typical inductive coding exercises where a code is manually assigned to an item by a coder, when we say an advice item is ``coded'' by a coder, we mean they used the coding tree, and the associated methodology assigned the resulting code. The coding software interface tool that we developed to ease coding and record results displayed: the coding tree (Fig.~\ref{fig:flowchart2}), code definitions (Fig.~\ref{box:tags}), detailed annotations for each question (App.~\ref{app:cheatsheet}), and two drop-down boxes where coders were asked to input the codes delivered through use of the coding tree. 
    
    \textbf{Coding tree methodology summary.} 
    In summary, an iterative inductive coding methodology was used both to derive codes and build the coding tree, and the coding tree was used to code the 1013 advice items (Section~\ref{sec:analysis}). While the coders were asked to follow the coding tree down to the leaf nodes and then enter the code delivered into the drop-down boxes, our software implementation did not prevent coders from immediately selecting a code, e.g., based on simply reading the advice item. (While coding reported here generally avoided use of such short-cut coding, this was not enforced by software. In retrospect, a preferred implementation would force coders to select \textit{yes}/\textit{no} answers until a code was automatically assigned to an advice item.)
    
    We note that many of the terms discussed in Section~\ref{sec:bestpractices} (e.g., categories from Section~\ref{sec:bestpractices}; the quality, commonality, and stipulation categories) are not represented in our codes. Where no extra context is provided about an individual advice item, and we use only the text of the advice (as was our case with analysis herein), it is difficult to know if an item belongs to any of these categories. For example, determining the quality of an advice item would require knowledge of how a community rates a practice, to know its commonality among practitioners requires knowledge of how frequently that advice is used, and to know if it is stipulated requires knowledge of how that practice is mandated in possibly widely varying real world environments. As such, terms like \textit{best practice}, \textit{common practice}, or \textit{regulation} are not used in our codes (Fig.~\ref{box:tags}). We have instead used codes that can be applied to advice items without requiring (unavailable) contextual information.
    
    A decision was made to use a single coder for the 1013-item coding exercise described in this paper, which was based on all test coders reaching a consensus on the final set of codes and questions in the coding tree (consistent with the methodology of, e.g., Huaman et al. \cite{Huaman2021}), the acceptably high level of agreement during test codings, and the work effort required to manually code (via the coding tree) 1013 items. Using a single coder is noted as a limitation of this work. Supplemental work (to be reported separately) will explore a second full coding of the 1013 items by at least one additional coder and pursue detailed explanations of any major deviations found. 
    
    As the advice items in the dataset are grouped by the 13 guidelines in the DCMS mapping document \cite{DCMS1}, all advice items in the coding of the full 1013 item set were randomly ordered to avoid bias from reading similar advice in repetition.
    
    \subsection{Advice Categorization by Lifecycle Phase}
    \label{sec:methodology/lifecycle}
    Separate from the inductive coding of Section~\ref{sec:methodology/flowchart}, the first coder assigned each actionable item (\textit{P3}--\textit{P6}) to a stage in the IoT lifecycle (Fig.~\ref{fig:lifecycle}) where the item could be best carried out (in the subjective opinion of the coder), independent of the codes defined and used in the coding tree. This was done by determining which stakeholder would be in a position (in our view) to carry out the item, and matching where this would appear to best occur in the lifecycle. This determination was based on which stakeholders \textit{could reasonably} execute a practice (within reason---an end-user given an API would not be likely to implement a best practice or fix a vulnerability), not necessarily the single stakeholder in the \textit{best/most effective position} to implement it, thus allowing for items to be associated with multiple stages. For example, Item \#191 \cite{GSMA4} states:
    
    \begin{quote}
        \textit{When a product is being developed it is often enabled with debugging and testing technologies to facilitate the engineering process. This is entirely normal. However, when a device is ready for production deployment, these technologies should be stripped from the production environment prior to the definition of the Approved Configuration.}
    \end{quote}
    
    \noindent This item could either be executed in the OS/App Development stage (1.2b) where code is stripped from software before completion, or during the Integration \& Pre-Configuration (1.3) stage where features may be disabled or left out of device integration. We considered only practices (being implicitly actionable) for this categorization, as without actionability, it is difficult to determine what steps would need to be taken and when (in the lifecycle) they would be executed.
    
    While an indication of where the advice item would be carried out in the lifecycle was included by many items in the advice statement itself (e.g., do not hard-code secret access codes for testing/debugging in software \cite{IOTSI2}), others required subjective judgement for placement (e.g., ``\textit{keep software updated}'' \cite{ETSI1} could be targeting the Creation phase or Usage phase depending on which stakeholder it implies should maintain software). If the item did not have an obvious or implied associated lifecycle phase, we categorized it as \textit{Unclear} (see Fig.~\ref{fig:lifecycle_actionable}).

    \subsection{Relationship to Security Principles}
    \label{sec:methodology/principles}
    We observed that many security advice items were rephrasings of established security principles. In our context of computer security, we define a \textit{principle} to be a generic (applying to many situations) rule shown through experience to improve security outcomes or reduce exposures, and a \textit{design principle} to be a subset specifically guiding the \textit{design} of a system. Other subsets may relate to other lifecycle phases. For context, note Saltzer and Schroeder \cite{Saltzer1975} define eight ``\textit{[...] useful principles that can guide the design and contribute to an implementation without security flaws}''. NIST \cite{NISTsp80027reva} notes ``\textit{the primary focus of these principles is the implementation of technical controls}'',\footnote{
        Examples of these principles (from the quote) are ``\textit{protect against all likely classes of `attacks'}'', ``\textit{use unique identities to ensure accountability}'', and ``\textit{limit or contain vulnerabilities}'' \cite{NISTsp80027reva}.
    } suggesting that security principles are appropriate targets to be implemented via practices. In our coding tree, both security principles and (more specifically) security design principles are leaf nodes.
    
    Fig.~\ref{fig:continuum} conveys our view of the relationship between security principles, practices, outcomes, and other terms. On one extreme (left) are concepts more focused on end-result (outcomes); on the other are the most actionable items that focus on mechanisms to reach outcomes, often specified in fine detail (standards). As Fig.~\ref{fig:continuum} indicates, some items on this continuum may serve as a guideline or requirement (Table~\ref{tab:qualifyingterms}). Ideally, in our view, what is imposed as requirements by governing bodies should be practices or standards (versus principles or outcomes), as requirements should be actionable so those subject to them have a clear understanding of how to follow them (cf. Section~\ref{sec:background/actionvsoutcome} for \textit{actionable}). This is represented on this continuum by the labeling of \textit{Practice} and \textit{Standard} as potential targets for requirements (denoted by the triangle).
    
    \subsection{Actual Use of Security Advice Coding Tree Methodology}
    We envision the methodology described in Section~\ref{sec:methodology} to be used primarily in two ways. The first is for measuring the actionability of existing advice as a means to establish a general view of the current state of IoT security advice, and to determine where advice fails to meet the needs of security practitioners. This is the primary focus of the coding exercise described in Section~\ref{sec:methodology}, and analyzed in Section~\ref{sec:analysis}. 
    
    The methodology can be used in a second way for the analysis of new IoT security advice, as a tool to assist advice authors in creating actionable advice. If advice authors themselves use the coding tree on their own advice items, they can differentiate actionable from non-actionable advice (among other more fine-grained characteristics of advice). After using the coding tree, security advice items that analysis tags with an undesirable code\footnote{\label{foot:undesirable}
        While we view the creation of actionable practices as a preferred objective for advice authors, and in this paper advocate the pursuit of actionable codes \textit{P4}--\textit{P6}, we acknowledge that creating actionable practices may not be the goal of all advice authors---some may intentionally craft non-actionable guidance in the form of Outcomes (\textit{O1a/b}), General Practices/General Policies (\textit{P2}), or Security Principles (\textit{N1}/\textit{N2}).  Fig.~\ref{box:tags} describes these categories. In any case, whether actionable or not, advice authors may use the SAcoding method to cross-check that their advice items match the categories that they intend.
    } can then be revisited by advice authors to revise, reword, and clarify the explanation of the advice. 
    
    Once an advice item is revised, the questions in the coding tree may yield different answers, giving advice authors feedback about whether their changes have had a positive impact on the actionability of their advice, or if it follows a path down the tree to a more desirable code. If the coding tree outputs an undesirable code (e.g., non-actionable), those giving the advice may be able to observe (from the coding tree) at which question the advice diverged from a path to a desired code. 
    
    For example: ``\textit{encrypt stored passwords}'' gives vague advice and is ambiguous on how to achieve encryption. Question 5 (from the coding tree, Fig.~\ref{fig:flowchart2}) sends this advice to \textit{Incompletely Specified Practice} (\textit{P1}), as it lacks actions to take. The advice item could be reworded to specify a particular encryption algorithm and mode. An accompanying document or note could also provide explicit references to aid implementation, thus now passing \textit{Q5} as actionable. 
    
\section{Empirical Analysis of IoT Security Advice Dataset}
\label{sec:analysis}

In this section we carry out our analysis of the 1013-item IoT security advice dataset, which for context, is the dataset of security advice from which the DCMS created their 13 guidelines \cite{DCMS2}. We coded the items in this collection using the methodology of Section~\ref{sec:methodology}. The primary goal of assigning each item to codes (and associated definitions) is to provide a general sense of how well existing advice dataset lists and literature specify practices (as opposed to advice \textit{positioned} as practices, but failing to be actionable, as required by our definition). Identifying where practices are carried out throughout the IoT lifecycle allows us to see which stakeholders are in the best position, or have the greatest number of items to address regarding contributing to overall device security.

    \subsection{Results of Coding}
    \label{sec:analysis/mainfindings}
    
\begin{figure}[hbt]
        \centering
        \includegraphics[width=0.85\textwidth]{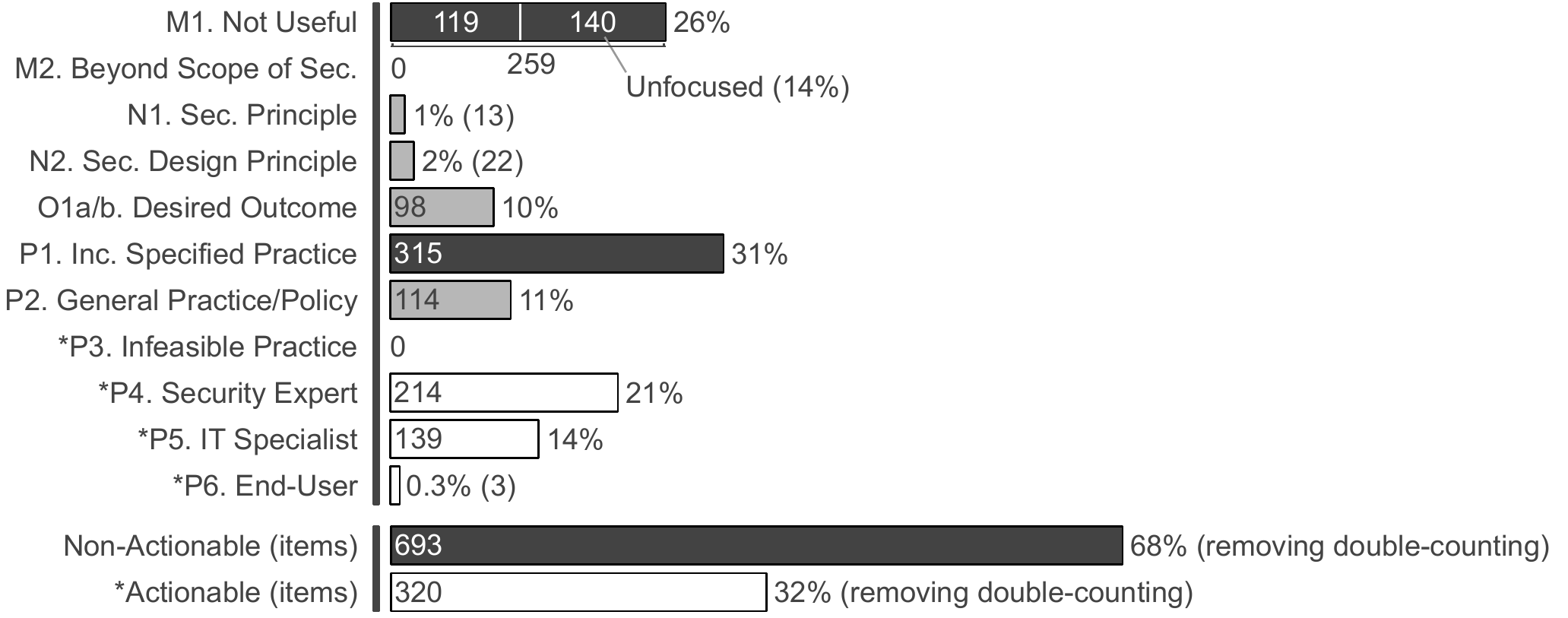}
        \caption{Number of items receiving each code, from coding of 1013-item advice dataset \cite{BellmanDataset}. Sum of counts in top portion exceeds 1013 as each item could be assigned one or two codes (first/second, per Section~\ref{sec:methodology/flowchart}). Removing resulting double-counting, the number of actionable items is $320/1013 = 32$\% (not: $214 + 139 + 3 = 356/1013 = 35.1$\%). Shading intensity follows the scheme of Fig.~\ref{fig:flowchart2}.}
        \label{fig:tagresults}
    \end{figure}

    Fig.~\ref{fig:tagresults} summarizes the distribution of codes given to all advice items in the DCMS 1013-item dataset, as coded by the first coder (C1). For distinguishing actionable vs. non-actionable advice (bottom of figure), if an item had two codes (first and second) and at least one was an actionable code (i.e., \textit{P3}--\textit{P6}), we declared the item actionable based on the reasoning that an argument could be made for its actionability. For example, if an item was coded both as an \textit{Incompletely Specified Practice} (\textit{P1}) and \textit{Specific Practice---Security Expert} (\textit{P4}), we declared the item actionable as \textit{P4} is defined to be actionable. As such, the sum of actionable and non-actionable items in Fig.~\ref{fig:tagresults} adds to 1013. 
    
    \raggedbottom
    
    The coding tree's software interface allowed a coder to designate whether an item was specific to IoT. None of the items in the dataset were designated in this way. While herein the coding tree is used to explore IoT security, we believe that its design and resulting structure apply more broadly, to analysis of security advice in general. Restated, the design intent is a generic (non IoT-specific) method.
    
    \subsection{Proportion of Non-Actionable Advice}
    \label{sec:analysis/actionability}
    Our analysis shows that organizations---often highly credible ones---are producing recommendations for manufacturers that are not, by our definitions and analysis, actionable, thus we believe at greater risk of being improperly (or not at all) executed or implemented. This low proportion of actionable practices (of the 1013-item set) is, in our view, a signal that the security community must significantly improve how we capture and state ``best practices'' if manufacturers are expected to follow recommendations.
    
    The methodology used declares any code after the \textit{yes} branch of $Q_5$ in the coding tree (\textit{P3}--\textit{P6}) to be actionable (per Section~\ref{sec:background/actionable}). As shown in Fig.~\ref{fig:tagresults}, in total 32\% (320/1013) of advice items were found to be actionable (at least one actionable code, per Section~\ref{sec:methodology/flowchart}). This includes the following tags (Fig.~\ref{fig:tagresults} caption explains over-counting): 
    
    \begin{itemize}
        \item 21\% of items (214/1013) had one tag (of possibly two) being \textit{P4}  (\textit{Specific Practice---Security Expert}); 
        \item 14\% of items (139/1013) had one tag (of possibly two) being \textit{P5}  (\textit{Specific Practice---IT Specialist}); and
        \item \textless1\% of items (3/1013) had one tag (of possibly two) being \textit{P6} (\textit{Specific Practice---End-User}). 
    \end{itemize}
    
    \noindent The \textit{Infeasible Practice} (\textit{P3}) code went unused; this was encouraging, suggesting that advice providers have an understanding of what sorts of practices are feasible (in both resources and knowledge) for their target audience. Similarly, the code \textit{Beyond the Scope of Security} (\textit{M2}) was also unused; however, this is arguably due to source documents being generally targeted at computer security.
    
    As Fig.~\ref{fig:tagresults} notes, 68\% of the advice was declared to be non-actionable. We expect that this significant majority of advice items may often be poorly implemented (or not at all, a failure in both cases)  despite advice recipients' best efforts to understand the advice. This does not imply that non-actionable advice is in all cases detrimental---outcomes, principles, and general practices still specify desirable end-results (outcomes) and generic goals. Actionability may not be essential in all use cases (cf. footnote~\ref{foot:undesirable}); however, our underlying assumption is that advice givers (for the advice datasets under discussion) intend to be offering advice positioned as best practices. We argue that actionability should be considered a high (if not the highest) priority among the characteristics of such security advice (see Section~\ref{sec:background/actionvsoutcome}).
    
    \subsection{`Not Useful' Advice}
    \label{sec:analysis/notuseful}
    During the iterative development of our coding tree methodology, we observed many advice items in the DCMS 1013-item dataset (described in Section~\ref{sec:background/dcms}) tended to not be actions to take, but descriptions of security techniques (e.g., a hardware security module, public-key encryption) or threats to a system (e.g., unused but accessible network ports), and offered no suggestion for any action to take or execute. Item \#387 \cite{IIC1} provides an example of this:
    
    \begin{quote}
        \textit{Network firewalls are message-oriented filtering gateways used extensively to segment IIoT [industrial IoT] systems. Most firewalls are Layer 2, 3 or 4 IP routers/message forwarders with sophisticated message filters. Firewalls may be deployed as either physical or virtual network devices. A firewall's filtering function examines every message received by the firewall. If the filter determines that the message agrees with the firewall's configured traffic policy, the message is passed to the firewall's router component to be forwarded.}
    \end{quote}
    
    \noindent One could make the argument that the description of a technique implies that the advice giver wants a follower to use the technique, but the italic text block above reads quite different from ``do this'' security advice and is lacking in actionable detail. As such, we consider advice of this nature to be not sufficient for a stakeholder to execute. 26\% (259/1013) of items were coded as \textit{Not Useful} (\textit{M1}); see Fig.~\ref{fig:tagresults} and description of \textit{M1} in Fig.~\ref{box:tags}. Note that \textit{M1} is also used for advice items that are judged to ``not make sense'' from a grammar or language perspective. 
    
    Similarly, individual advice ``sub-items'' are commonly given in rapid succession within a single advice item (which may take the form of several sentences or a paragraph). As a sub-category of the \textit{Not Useful} (\textit{M1}) code, we added an \textit{Unfocused} supplementary code for coders to use when they find multiple sub-items within one item (represented as the left sub-bar of the \textit{Not Useful} code in Fig.~\ref{fig:tagresults}). For example, Item \#84 \cite{BITAG1} is, in our view, an example of this: 
    
    \begin{quote}
        \textit{IoT Devices Should Follow Security \& Cryptography Best Practices. 
        [1] BITAG recommends that IoT device manufacturers secure communications using Transport Layer Security (TLS) or Lightweight Cryptography (LWC). Some devices can perform symmetric key encryption in near-real time. In addition, Lightweight Cryptography (LWC) provides additional options for securing traffic to and from resource constrained devices. [2] If devices rely on a public key infrastructure (PKI), then an authorized entity must be able to revoke certificates when they become compromised, as web browsers and PC operating systems do. Cloud services can strengthen the integrity of certificates issued by certificate authorities through, for example, participating in Certificate Transparency. [3] Finally, manufacturers should take care to avoid encryption methods, protocols, and key sizes with known weaknesses. [4] Vendors who rely on cloud-hosted support for IoT devices should configure their servers to follow best practices, such as configuring the TLS implementation to only accept the latest TLS protocol versions.}
    \end{quote}
    
    \noindent This advice item jumps across four topics (we inserted the numbers for exposition): (1) the use of TLS or lightweight cryptography, (2) certificate revocation, (3) avoiding weak or vulnerable key sizes, and (4) avoiding outdated TLS versions. Trying to successfully code advice such as this (i.e., as a coder) was a challenge, as different sub-items could be coded differently. 
    
    We found that advice items with a longer word length were often unfocused in this way. In total, 54\% (140/259) of the items that were tagged with \textit{Not Useful} (\textit{M1}) codes, or 13.8\% (140/1013) of all items were coded as \textit{Unfocused}, implying that in the judgement of the coder they contained multiple distinct topics within the advice item (similar to the above example). Had we extracted each sub-item from the original dataset (making them more narrowly focused, but as a result, potentially removing them from surrounding context), these may have been coded differently, suggesting a limitation to this portion of our work.
    
    \subsection{Associating Advice Items with IoT Lifecycle Stages}
    \label{sec:analysis/lifecycle}
    
\begin{figure}[t]
        \centering
        \includegraphics[width=0.80\textwidth]{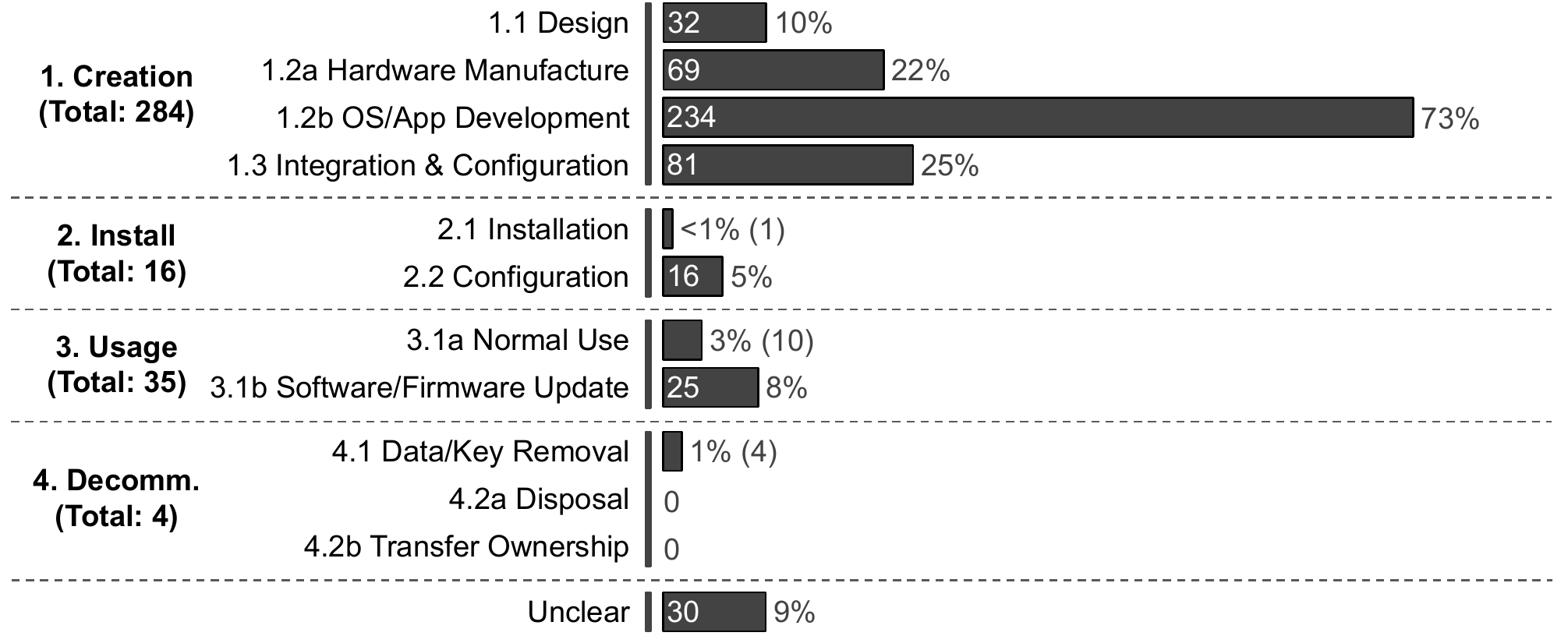}
        \caption{Number of actionable practices that we declared as suitable to implement at each lifecycle phase (Fig.~\ref{fig:lifecycle}). Total of all numbers and percentages exceed actionable practice total (320), phase totals, and 100\% as practices may be suitable to implement in multiple stages. Percentages are proportion of 320 actionable practices.}
        \label{fig:lifecycle_actionable}
    \end{figure}

    Fig.~\ref{fig:lifecycle_actionable} shows the results of one coder's manual association of each of the 320 actionable practices (Fig.~\ref{fig:tagresults}) with one or more of the lifecycle stages. All lifecycle phases associated with each practice (Section~\ref{sec:methodology/lifecycle}) were combined to yield the count shown for each bar. For example, if an item was coded as taking place in both the OS/App Development (1.2b) and Design (1.1) stages, each of these counts were increased by one (1).
    
    \raggedbottom
    
    The practice distribution among phases reveals important information about the overall execution of best practices: 89\% (284/320) of practices that were deemed actionable could be implemented in at least one lifecycle phase within the manufacturer's control (i.e., the Creation phase), i.e., the designers and manufacturers are in a position to implement them. As a subset of this, the OS/app developers alone (Phase 1.2b, Fig.~\ref{fig:lifecycle}) are in a position to implement 73\% (234/320) of practices. This follows from many advice items being software-related, thus suitable for implementation by one or more of several stakeholders involved in software development, before the product is in end-user hands. 
    
    While this finding may seem self-evident, it draws focus to the importance of attention to security during the product (device) Creation phase, and in particular, the importance of IoT security advice being implementable (and understandable) by IoT device manufacturers and their software development partners.

\section{Related Work}

Explicit formal definitions for the term \textit{best practice} are rare in the security literature. Literature about the nature and definition of security best practices (as opposed to examples of best practices) is discussed in Section~\ref{sec:bestpractices}. 
In the remainder of this section we discuss related work on establishing security practices (for IoT and other areas). Tschofenig and Baccelli \cite{Tschofenig2019} discuss efforts by The European Union Agency for Cybersecurity (ENISA) and the IETF to provide recommendations and specifications on IoT security. They categorize technical and organizational areas to be considered for the secure development and use of IoT devices. Moore et al. \cite{Moore2017} pursue specific best practices for IoT, specifically regarding network-based attacks. Based on our definitions herein, most of their advice items are not actionable (thereby not what we consider to be practices). Dingman et al. \cite{Dingman2018} examined six sets of IoT security advice and looked to determine whether three large-scale security events may have been averted if their collected security advice was followed. 

Alrawi et al. \cite{Alrawi2019} analyze and systematize work on home-based IoT security and propose a methodology for evaluating the security of home-based systems. They note that best practices are ``readily available'', but provide neither definitions nor specific references. Assal and Chiasson \cite{Assal2018} note that for software development security practices, technical detail is inconsistent, and find common security advice is often not followed. Redmiles et al. \cite{Redmiles2020} analyze security advice for end-users and find through a user-study that most advice is viewed (by users) as being actionable, but it is unclear to them which of the actionable advice is the most important to follow (i.e., which advice to prioritize from the set). Acar et al. \cite{Acar2017} analyze software development security advice and find it is often inadequate for software developer needs, and lacking resources (e.g., implementation examples, tutorials) to help them understand the guidance. 

A number of government and industrial agencies provide security advice for IoT, for both manufacturers and groups looking to acquire IoT devices for their organizations. ENISA \cite{ENISA1} published an expansive document about IoT security. This includes a substantial set of security recommendations, but also useful contextual and informative sections including (to single out a select few) the document's target audience (cf. Section~\ref{sec:background/actionvsoutcome}), an overview of what IoT is and the relevant components, threat and risk analyses, and technical measures for executing the advice (these measures appear to be positioned as technical steps to complement other security advice). 

ETSI \cite{ETSI2020} (cf. Section~\ref{sec:background/dcms}) provides a series of baseline requirements for IoT security. These requirements use the 13 DCMS guidelines \cite{DCMS2} as general topic headers (adding a new one of their own), but provide more detail about technical steps to be taken. To supplement the baseline requirements document, ETSI provides a document describing how to confirm conformance with the advice therein, noting that the advice in the support document is independent of an assurance scheme \cite{ETSI2021}. Assurance is historically associated with products for governmental use \cite[Chapters 18--21]{Bishop2003} \cite[Chapter 13]{Gollmann2011}, but is typically considered too expensive or otherwise unsuitable to the consumer space. For DCMS documents \cite{DCMS1, DCMS2, IoTSecMap} used in this paper, see Section~\ref{sec:background/dcms}. 

NIST published three documents surrounding the interaction between US federal government agencies and IoT manufacturers \cite{NISTIoTNews}. Two of these NIST documents \cite{NISTIR8259b,NISTIR8259c} aim to assist IoT manufacturers to produce secure devices specifically for use in the US federal government by offering technical and non-technical baseline guidance. One of the NIST documents \cite{NISTsp800213} is intended to help government agencies learn what features or characteristics they should seek when procuring IoT devices. 

RFC 8576 \cite{Garcia-Morchon2019} proposes a generic lifecycle model of an IoT device, presented as a simplified model. Other descriptions of lifecycles may include the key functional components that describe its primary function (versus the entirety of its life), e.g., Alrawi et al.'s IoT malware lifecycle components \cite{Alrawi2021}. NIST's SP 800-27 \cite{NISTsp80027reva} outlines five major general computer and IT system lifecycle phases and suggests 33 IT security principles. While developed independently, our lifecycle of Section~\ref{sec:background/lifecycle} unsurprisingly has similarities, e.g., design/development, primary usage, and disposal/end-of-life phases. The NIST SP suggests that many individual principles are vital to positive security outcomes across multiple phases, implying there are important principles for phases other than the design phase. NIST SP 800-160 \cite{NIST2} outlines a taxonomy of 32 security design principles covering three areas of systems security: \textit{security architecture and design}, \textit{security capability and intrinsic behaviors}, and \textit{lifecycle security}; the latter two are not specific to the design phase.

Morgner et al. \cite{Morgner2018} explore the relationship between efforts in formal IoT technical standards and the (unfortunate) reality of the economics of IoT security and its implications for the general security of manufacturers. 

\section{Concluding Remarks}

The basic concept of best practices is familiar to non-experts. Our analysis of a wide selection of security advice found conflation of the ideas of security goals (outcomes) and the steps or methods by which they may be reached (practices). We highlighted an important characteristic for security advice: whether or not it is actionable. We offer uniform, consistent terminology (Section~\ref{sec:bestpractices}) that characterizes and separates concepts. This allowed systematic exploration that began with generic discussion and analytic classification and was cross-checked through specific focus on consumer IoT devices and their lifecycle. 

A main contribution of this paper is the development of the security advice coding tree methodology and its use to systematically analyze a large collection of IoT security advice. In particular, we examined how actionable current advice is (we use the DCMS 1013-item dataset as representative of current IoT security advice), and what advice characteristics (i.e., corresponding to the codes of Fig.~\ref{box:tags}) emerge from this dataset. Our main focus has been the DCMS 1013-item IoT security advice dataset, which itself originates from other organizations. 

For our analysis, iterative inductive coding was used to create a codebook that represents the characteristics of security advice (e.g., whether they are objectives to reach or practices to follow). To more objectively assign a code to each item in the dataset, we designed a coding tree. We suggest that IoT security advice-giving organizations consider using the coding tree and methodology of Section~\ref{sec:methodology} to measure whether potential advice is actionable, and take steps to improve the advice's actionability, unless their explicit goal is to target, e.g., security principles or outcomes to reach. 

From our analysis of 1013 advice items from industrial, governmental, and academic sources, we were surprised to find that the majority of advice items are not actionable practices, but rather, what we deem to be non-actionable advice. Among the practices we identified as actionable, 73\% are suitable to implement in the OS/App Development lifecycle phase of an IoT device (Section~\ref{sec:analysis/lifecycle}), thus by the product manufacturer and its software development partners. It is generally recognized that poor security practices early in the lifecycle accrue what we might call a \textit{security debt}, with negative consequences in later phases (analogous to \textit{tech debt} where technical shortcuts during development incur later costs \cite{Kruchten2012}); from this, our results herein highlight the fundamental role of pre-deployment stakeholders to underpin security for aspects that they alone are in a position to control. 

As the Internet of Computers has grown into the Internet of Things, an old problem remains: how to ensure that security best practices are followed. An open question is whether the research community can find ways to help advice-givers (including governments) to compile more effective guidance, and have manufacturers embrace and execute advice given. Our work argues that currently, even in the security and technology communities (not to mention the general public), ambiguity surrounds the language of technical \textit{best practices}---such that arguably, the term does as much harm to the security community as good. One hypothetical path forward (to provoke thoughts, more than as a practical suggestion) is to seek agreement within the technical security community that the term itself is vague and nebulous, and its use should be boycotted. Another path forward is to work towards consensus on definitions (as we pursue herein).

We suggest that organizations proposing and endorsing ``best practice'' advice have a clear idea of whether they are recommending practices, specifying baseline security requirements, or simply offering advice about good principles to think about. If the goal is that relevant stakeholders adopt and implement specific practices aiming to reduce security exposures, we believe it is imperative that (actionable) best practices be identified and clearly stated, versus vague outcomes---lest the target stakeholders be unable to map advice to a concrete practice, even if so motivated. In summary: if security experts do not find guidelines clearly actionable, we should not expect (security non-expert) manufacturers to magically find a way to adopt and implement the advice. The economic motivation of manufacturers \cite{Morgner2018} (keeping in mind markets for lemons \cite{Akerlof1970}), their poor track record in IoT security, and lack of accountability for vulnerabilities, point to a worrisome future. We hope that our work is a step towards improving the efficacy of advice on best practices.

\begin{acks}
    We thank anonymous referees for helpful comments. This work was supported by the Natural Sciences and Engineering Research Council of Canada (NSERC), which is acknowledged for Discovery Grants to the first and third authors, and for funding the third author as Canada Research Chair in Authentication and Computer Security.
\end{acks}

\bibliographystyle{ACM-Reference-Format}
\bibliography{bib}

\appendix

\section{Detailed Annotations for Coding Tree Questions}
\label{app:cheatsheet}

Items Q1--Q11 below and the adjacent text are the detailed annotations for the coding tree questions, noted in Section~\ref{sec:methodology}. These were consulted if a coder was unclear what a question was asking or for further details, to answer a  question for an advice item being coded.

\begin{itemize}
\item [Q1.]
\textit{Is the advice conveyed in unambiguous language, and relatively focused?}

Does the advice make sense from a language perspective (e.g., it is a sentence that you can read and makes sense), unambiguous (i.e., you can tell what they are trying to convey from a language perspective, not technical), and not multiple items grouped into one piece of advice? Is the advice focused on one topic, whether it is a step to take, an outcome to achieve, or security principle? If the advice seems to have multiple topics being discussed or has multiple outcomes it wants an implementer to reach, this would be considered unfocused.

\item [Q2.]
\textit{Is it arguably helpful for security?}

Is the advice arguably useful for pursuing security in some way? Does it seem like it will help improve security outcomes rather than processes unrelated to security?
 
\item [Q3.]
\textit{Is it focused more on a desired outcome than how to achieve it?}

Is the advice a high-level outcome rather than some method (or meta-outcome) for how to achieve an outcome? E.g., \textit{data is secured in transit} would be an outcome because it is a desired goal or state, whereas \textit{encrypt data in transit} is not because it explains a method for achieving that outcome (in this case, encryption). \textit{Encryption} may be considered a meta-outcome, as it is not meaningful to the end-user's ultimate goal of protected data.

\item [Q4.]
\textit{Does it suggest a security technique, mechanism, software tool, or specific rule?}

Is the item a method used in achieving/following the advice? E.g., \textit{encryption} or \textit{replacing a password with black dots} are techniques/mechanisms, but \textit{secure data} or \textit{making the password unreadable} are not. An example of a specific rule: \textit{no hard-coded credentials}---this is a rule that is fairly specific as to its goal and would be followed like a practice, but not necessarily with actionable steps.

\item [Q5.]    
\textit{Does it describe or imply steps or explicit actions to take?}

Does the advice suggest actionable technical steps (one or more) that suffice to follow the advice? It has sufficient detail to suggest a step/action to take.
    
\textbf{Actionable}: Involving a known, unambiguous sequence of steps, whose means of execution is generally understood.

\item [Q6.]    
\textit{Is it viable to accomplish with reasonable resources?}

Could the advice item be followed with an acceptable cost?. E.g., the advice would not take years to follow, or have cost out of line with the anticipated benefit. 

\item [Q7.]    
\textit{Is it intended that the end-user carry this advice out?}

Does the item suggest that the end-user will be responsible for carrying out this practice? Note that end-users first interact with devices after the Creation phase.

\item [Q8.]    
\textit{Is it intended that a security expert carry this item out?}

Does following this advice require an expert understanding of security and security implementation in order to properly follow the advice? Someone following this advice item would have to be an expert in security to be able to understand it and successfully follow it, or be capable of extracting actionable steps from an otherwise non-actionable item based on their experience.

\item [Q9.]    
\textit{Is it a general policy, general practice, or general procedure?}

Is the item a security policy (general rule) to improve security, but is not explicit about what technical means is used? These are less actionable (akin to incompletely specified practices---see definition in Q5), and are not technically explicit. A general policy often has more emphasis on what is (dis)allowed (or may be a general rule closely related to a desired outcome), rather than on how to achieve it.

\item [Q10.]    
\textit{Is it a broad approach or security property?}

Is the item a general way or general strategy, or a property that would improve security? A security property is a characteristic or attribute of a system related to security. E.g., an \textit{open design}.

\item [Q11.]    
\textit{Does it relate to a principle in the design?}

Some principles relate to the core design phase of the product/system rather than later lifecycle phases.

\end{itemize}

\end{document}